\documentclass[preprint,aps,showpacs]{revtex4}
\usepackage[latin9]{luainputenc}
\usepackage{float}
\usepackage{amsmath}
\usepackage{amssymb}
\usepackage{graphicx}
\usepackage{setspace}

\makeatletter

\@ifundefined{textcolor}{}
{%
 \definecolor{BLACK}{gray}{0}
 \definecolor{WHITE}{gray}{1}
 \definecolor{RED}{rgb}{1,0,0}
 \definecolor{GREEN}{rgb}{0,1,0}
 \definecolor{BLUE}{rgb}{0,0,1}
 \definecolor{CYAN}{cmyk}{1,0,0,0}
 \definecolor{MAGENTA}{cmyk}{0,1,0,0}
 \definecolor{YELLOW}{cmyk}{0,0,1,0}
}

\usepackage{amsfonts}
\usepackage{bm}
\usepackage[dvips]{epsfig}
\usepackage[dvips]{graphicx}
\usepackage{float}
\usepackage{dcolumn}
\usepackage{ifpdf}
\usepackage{multirow}

\newcommand{\sech}{\mathrm{sech}}
\newcommand{\be}{\begin{equation}}
\newcommand{\ee}{\end{equation}}
\newcommand{\bes}{\begin{subequations}}
\newcommand{\ees}{\end{subequations}}
\newcommand{\ben}{\begin{eqnarray}}
\newcommand{\een}{\end{eqnarray}}

\makeatother

\begin{document}

\title{Boundary scattering in the $\phi^{6}$ model}

\author{Fred C. Lima$^{1}$, Fabiano C. Simas$^{2}$, K. Z. Nobrega$^{3}$, Adalto R. Gomes$^{1}$
}
\email{fredfjcl@gmail.com,simasfc@gmail.com,bzuza1@yahoo.com.br,argomes.ufma@gmail.com}


\affiliation{
$^{1}$ Departamento de F\'isica, Universidade Federal do Maranh\~ao
(UFMA) \\
 Campus Universit\'ario do Bacanga, 65085-580, S\~ao Lu\'is, Maranh\~ao,
Brazil
 $^{2}$ Centro de Ci\^encias Agr\'arias e Ambientais-CCAA, Universidade
Federal do Maranh\~ao (UFMA), 65500-000, Chapadinha, Maranh\~ao, Brazil\\
$^{3}$ Departamento de Eletro-Eletr\^onica, Instituto Federal de Educa\c c\^ao,
Ci\^encia e Tecnologia do Maranh\~ao (IFMA), Campus Monte Castelo, 65030-005,
S\~ao Lu\'is, Maranh\~ao, Brazil
 }
\begin{abstract}
We study the non-integrable $\phi^{6}$ model on the half-line. The model has two topological sectors. We chose solutions from just one topological sector to fix the initial conditions. The scalar field satisfies a Neumann boundary condition $\phi_{x}\left(0,t\right)=H$.   We study the scattering of a kink (antikinks) with all possible regular and stable boundaries. When $H=0$ the results are the same observed for scattering for the same model in the full line. With the increasing of $H$, sensible modifications appear in the dynamics with of the defect with several possibilities for the output depending on the initial velocity and the boundary.  Our results are confronted with the topological structure and linear stability analysis of kink, antikink and boundary solutions. 
\end{abstract}

\pacs{11.10.Lm, 11.27.+d, 98.80.Cq}

\maketitle

\section{ Introduction }

Solitary waves/solitons have a large number of realizations in several areas of physics \cite{dauxois,weinberg}. Despite simpler, models in $(1,1)$ dimensions also have applicability in physics at all scales, from dark and bright solitons in cigar-shaped Bose-Einstein condensate \cite{be} to high energy physics \cite{vachaspati}. The kink (antikink) is the simplest solution with solitary character.

The solitary waves in integrable systems are characterized by a very simple behavior in kink-antikink collisions, with at most a phase shift in the field. Only solutions with this characteristic are defined as solitons in the mathematical literature, but in this paper we make no such restriction. The investigation of kink/antikink scattering in nonintegrable systems have since a long time showed some surprising results such as bion and two-bounce solutions.  Several classes of nonintegrable models have already been studied. One can cite the polynomials $\phi^4$ \cite{s,m,kk_nonintegr1,kk_nonintegr2,bk,kk_nonintegr3,kk_nonintegr4}, $\phi^6$ \cite{kk_nonintegr5,kk_nonintegr6},  $\phi^8$ \cite{kk_nonintegr10}, polynomials with extensions \cite{dr} or with an external perturbation \cite{rs} and nonpolynomials \cite{cps,pk,cp,gk,bbg,kk_nonintegr11,sgn}. Interesting scattering structures where observed with multikinks \cite{kk_nonintegr8,kk_nonintegr9,gma,kk_nonintegr12} and with models with two scalar fields \cite{rom2,alon1,alon2,alon3}.  Despite the existence of analytical results for some classes of potentials \cite{kk_nonintegr13}, the study of kink scattering is mostly numerical. 
 
As a first example, the  nonintegrable $\phi^{4}$ model presents a
rich structure, which depends crucially in their initial velocity $v_i$ \cite{kk_nonintegr3}.
For large initial velocity, above a critical
value $v_{c}$, the pair $K\bar{K}$ recedes from each other, i.e,
the kinks always escape to infinity after one collision. On the other
hand, for initial velocities bellow $v_{c}$, the structure of the collision
is far more complex. In this region, the kink and antikink capture
one another, forming the bion state. For $v_i\lesssim v_c$ there is the possibility of scattering after a two-bouncing process, where the pair is again able to escape to infinity after colliding twice \cite{kk_nonintegr2, kk_nonintegr3,kk_nonintegr4,s,m,bk}. Each two-bounce process can be identified as corresponding to a particular two-bounce window. Such windows in velocity are characterized by widths that are reduced with the grow of $v_i$ and accumulate around $v_i=v_c$. Near to each two-bounce windows usually there is a sequence of three-bounce windows. This strucure is reproduced in a fractal pattern \cite{kk_nonintegr3}.

According to Campbell, Schonfeld and Wingate (CSW)  \cite{kk_nonintegr2}, a collision presenting two-bounce is described by a resonance effect  between the zero mode and the vibrational mode of the kink. Firstly there is the transferring of energy from the translational mode to the internal (vibrational) one.
In the sequel, the energy is transferred
back to the zero mode and the kink-antikink pair is separated from their
mutual attraction. 

A counterexample
of the CSW mechanism was found in Ref. \cite{kk_nonintegr5} for the $\phi^{6}$ model. There it was shown
that certain antikink-kink collisions exhibit resonant scattering, even
in the absence of a vibrational mode for one kink. This surprising result was related to the existence of a bound state produced by the antikink-kink pair \cite{kk_nonintegr5}. A second counterexample was presented in Ref. \cite{kk_nonintegr11}, where some of us
 showed the total suppression of two-bounce windows, even with the presence
of more than one vibrational mode.

The study of nonlinear field theories with boundary have a long tradition in integrable systems, mostly connected to the sine-Gordon model \cite{integr1,integr2,integr3,integr4,integr5,integr6}.
One object of investigation is to find suitable boundary conditions compatible with
integrability \cite{gz}. Supersymmetric theories on the half-line keeping integrability were also considered \cite{susy1,susy2,susy3,susy4}. One interesting aspect that deserves to be more investigated is the consequence of nonintegrability in the interaction of a kink or antikink with a boundary. In this line there are few examples of previous investigation for the $\phi^4$ \cite{dorey1} and sine-Gordon \cite{dorey2} models.

The kink can be aplied in scenarions with large symmetry in high energy physics. For instance, the collision of relativistic bubbles can be treated as planar walls and described as kink scattering in $(1,1)$ dimensions \cite{bubble,kk_nonintegr7,gib}. Motivated by the undetection of topological defects expected to be produced at a large rate via the Kibble mechanism \cite{kibble} in the early universe, the authors of the Ref. \cite{rom5} investigated the $\phi^6$ model under a generic perturbation. They showed that, due to the counterintuitive negative pressure effect, any small perturbation could trigger a chain reaction to influence the stability of a system of kinks. In this way the negative pressure can be the cause of the vanishing of domain walls in some models.

 In condensed matter, the kink in an nonintegrable model was proposed theoretically in buckled graphene nanoribbon \cite{graph1, graph2}. The possibility of existence of the  negative radiation pressure effect in buckled graphene was considered in the Ref. \cite{rom1}, based in the existence of such effect in a kink the $\phi^4$ model \cite{rom4}. The connection of solitons with conducting polymers is an active area of research, and some interesting reviews can be found Refs. \cite{hks, lu} (for a  a more gentle introduction to the subject, see the Ref. \cite{rc}). Kinks have an important role for describing the intriguing metalic properties of heavily doped {\it trans}- polyacetilene.  In the Su-Schrieffer-Heeger (SSH) model \cite{ssh}, the kink/antikink excitations separate regions of two different classes of dimerization and satisfy the integrable sine-Gordon equation. These defects are spinless and are formed upon charge transfer from the dopant to the polymer. The transition to a metallic state was explained by Kivelson and Heeger \cite{kh} as a crossover from a soliton lattice to a polaron lattice.  Recently the interaction between the kink-antikink pair in photoexcited {\it trans}- polyacetilene was studied using ab-initio
excited state dynamics  \cite{poly}. After excited, the atoms evolve using an hybrid time-dependent density functional theory (TD-DFT). For $T=0$ a soliton/antisoliton nucleates and pass through each other, as expected by a sine-Gordon model. For $T\neq0$ a surprising result appears, with the solitons scattering after bouncing twice, a characteristic of a nonintegrable model. In an ideal conjugated polymer, a soliton is free to move because the total energy is independent on the position of the soliton. For finite chains, however, end effects push the soliton to the center of the chain. In the SSH theory, and for enlarged dimerization parameters,  a linear term is introduced in the potential leading to these repulsive effects of the border \cite{su-s, helm}. An extension of the SSH model to include a third-neighbor interaction \cite{pyc} shows that all conformal excitations in both {\it cis-} and {\it trans-}polyacetylene are repelled from the chain ends.

In this paper, we will discuss the process of collision of antikinks and kinks in the $\phi^{6}$
model with a Neumann boundary condition. Since the antikink (kink) $\phi^6$ has no associated vibrational state, our main motivation was to investigate if the boundary condition is able to produce bounce windows. Our numerical analysis shows that this is the case for antikink-boundary scattering, with a very intricate structure emerging from the scattering process. The kink-boundary scattering leads to different results depending on the type of boundary to be scattered, with the most notable aspect the possibility of he changing of the topological sector and the nature of the boundary due to the scattering. Anoher motivation was to investigate if the boundary has a repulsive character to the scattering deffect. We found that this is true in most cases, but for some configurations the defect can be ``trapped" around or even annihilated by the boundary.

In the next section we review the first-order formalism for obtaining static solutions and stability analysis. The  Sect. III reviews some known results of solutions and stability analysis for the $\phi^{6}$ model in the full line. In the Sect. IV  we present solutions and stability analysis in the half line. The numerical analysis of antikink boundary scattering is presented in the Sect. V.  Corresponding results for kink boundary scattering is presented in the Sect. VI. We conclude in the Sect. VII.

\section{BPS states}

We consider the following action with standard dynamics
\begin{eqnarray}
S=\int{dxdt\bigg(\frac{1}{2}\partial_{\mu}\phi\partial^{\mu}\phi-V(\phi)\bigg)}.
\end{eqnarray}
The equation of motion is given by
\begin{eqnarray}
\frac{\partial^{2}\phi}{\partial t^{2}}-\frac{\partial^{2}\phi}{\partial x^{2}}+V_{\phi}=0,\label{eom}
\end{eqnarray}
where $V_{\phi}=dV/d\phi$. If the potential is 
\be
\label{VW}
V(\phi)=\frac12 W_{\phi}^{2},
\ee
the solutions of the first-order equation
\begin{eqnarray}
\label{1order}
\frac{d\phi}{dx}=\pm W_{\phi},\label{primeira ordem}
\end{eqnarray}
are also static solutions of the second-order equation of motion. The defects
formed with this prescription minimize energy, connect adjacent minima
of the potential and are known as BPS defects \cite{bps1,bps2}.
Stability analysis around a static solution $\phi_{S}(x)$ considers 
$\phi(x,t)=\phi_{S}(x)+\eta(x)e^{i\omega t}$ 
and the first-order corrections in the equation of motion. This results in a Schr\"odinger-like eigenvalue equation 
\begin{eqnarray}
-\frac{d^{2}\eta(x)}{dx^{2}}+U(x)\eta(x)=\omega^{2}\eta(x),
\end{eqnarray}
where the potential is
\begin{eqnarray}
U(x)=\frac{d^{2}V(\phi)}{d\phi^{2}}\bigg|_{\phi=\phi_{S}}.
\end{eqnarray}
The Schr\"odinger-like equation can be rewritten as
${\cal H}\eta(x)=\omega^{2}\eta(x)$. That is, linear stability is assured
if the Hamiltonian ${\cal H}$ is positive definite.

Now we consider the model on the half-line $-\infty<x<0$. For this
we consider the action
\begin{equation}
S=\int_{-\infty}^{0}dx\int dt\biggl\{\frac{1}{2}\partial_{\mu}\phi\partial^{\mu}\phi-V(\phi)+\delta(x)H\phi\biggr\}.
\end{equation}
Varying the action with respect to $\phi$ leads to an equation of motion
for $x\neq0$ identical to Eq.(\ref{eom}) obtained previously for
the full-line. In addition, when extended and considered around $x=0$,
the action gives the integrable Neumann condition 
\be \frac{d\phi}{dx}\bigg|_{x=0}=H.\label{neumann}
\ee
Static solutions have energies given by
\be E=\int_{-\infty}^0
\biggl{[} \frac12 \biggl( \frac{d\phi}{dx} \biggr)^2 + V(\phi)
\biggr{]}dx - H \phi(x=0), \ee
or, by Eqs. (\ref{VW}) and (\ref{1order}),
\be 
\label{Ephi}
E[\phi]=\pm W\big|_{\phi(-\infty)}^{\phi(0)}-H\phi\left(0,t\right).
\ee

	\section{The $\phi^6$ model on the full line}

\label{secphi6}

We consider the model given by 
\begin{eqnarray}
V(\phi)=\frac{1}{2}\phi^{2}(1-\phi^{2})^{2}.
\end{eqnarray}
The associated superpotential  is $W(\phi)=\phi^2/2 - \phi^4/4$. The first-order equations are then
\begin{eqnarray}
\label{bps1}
\frac{d\phi}{dx}&=& \phi-\phi^3,\\
\label{bps2}
\frac{d\phi}{dx}&=& -(\phi-\phi^3).
\end{eqnarray}
Note that the potential has vacua given by $\{-1,0,1\}$. The kink connecting the minima
$\{0,1\}$ is given by \cite{kk_nonintegr5}
\begin{eqnarray}
\Phi_{1+}(x)=\sqrt{\frac{1+\tanh(x)}{2}}.
\label{Phi1+}
\end{eqnarray}
The antikink connecting the minima $\{1,0\}$ is given by \cite{kk_nonintegr5}
\begin{eqnarray}
\label{Phi2+}
\Phi_{2+}(x)=\sqrt{\frac{1-\tanh(x)}{2}}.
\end{eqnarray}
\begin{figure}
\includegraphics[width=8cm, height=4cm]{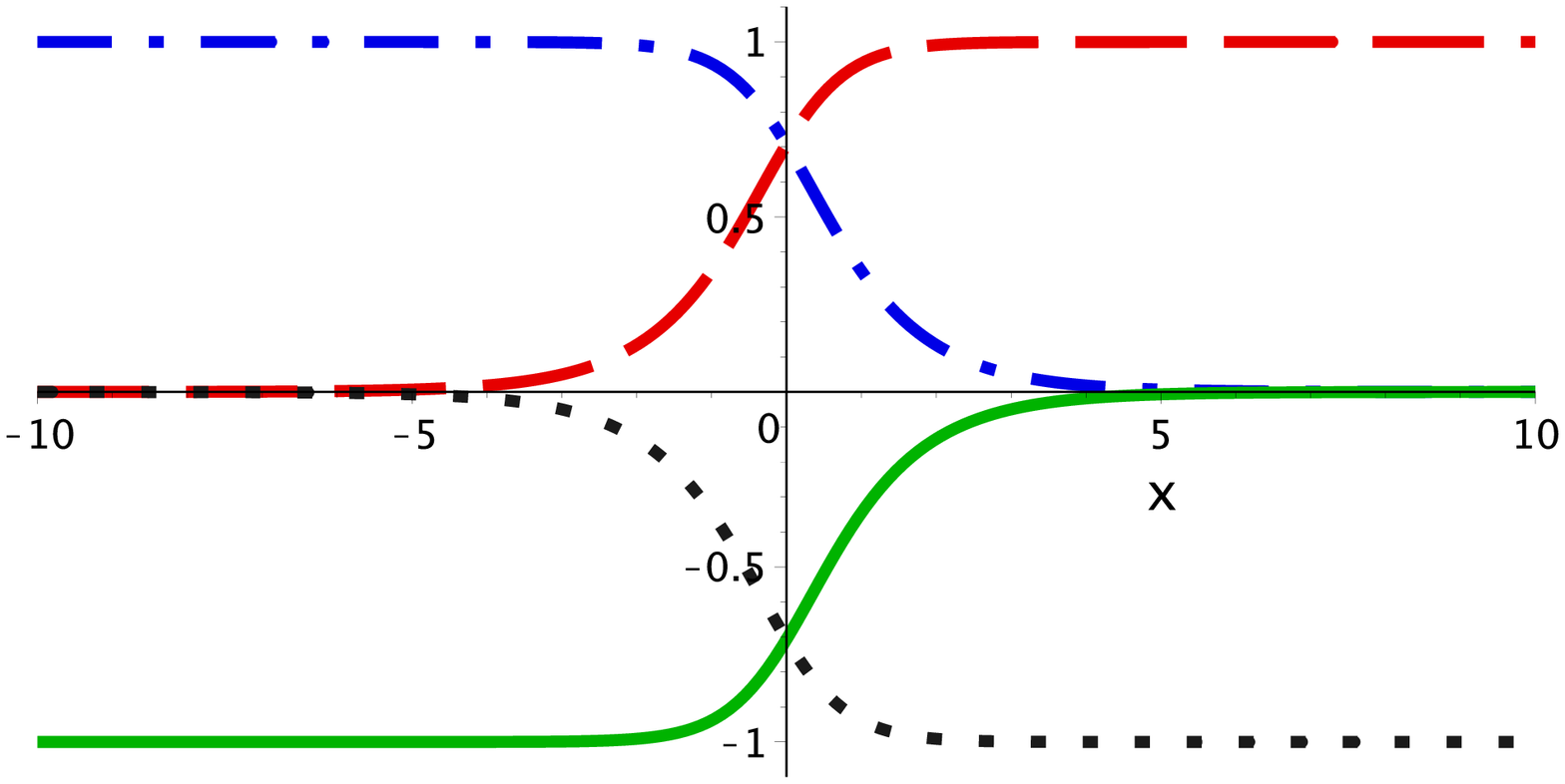} 
\includegraphics[width=8cm, height=4cm]{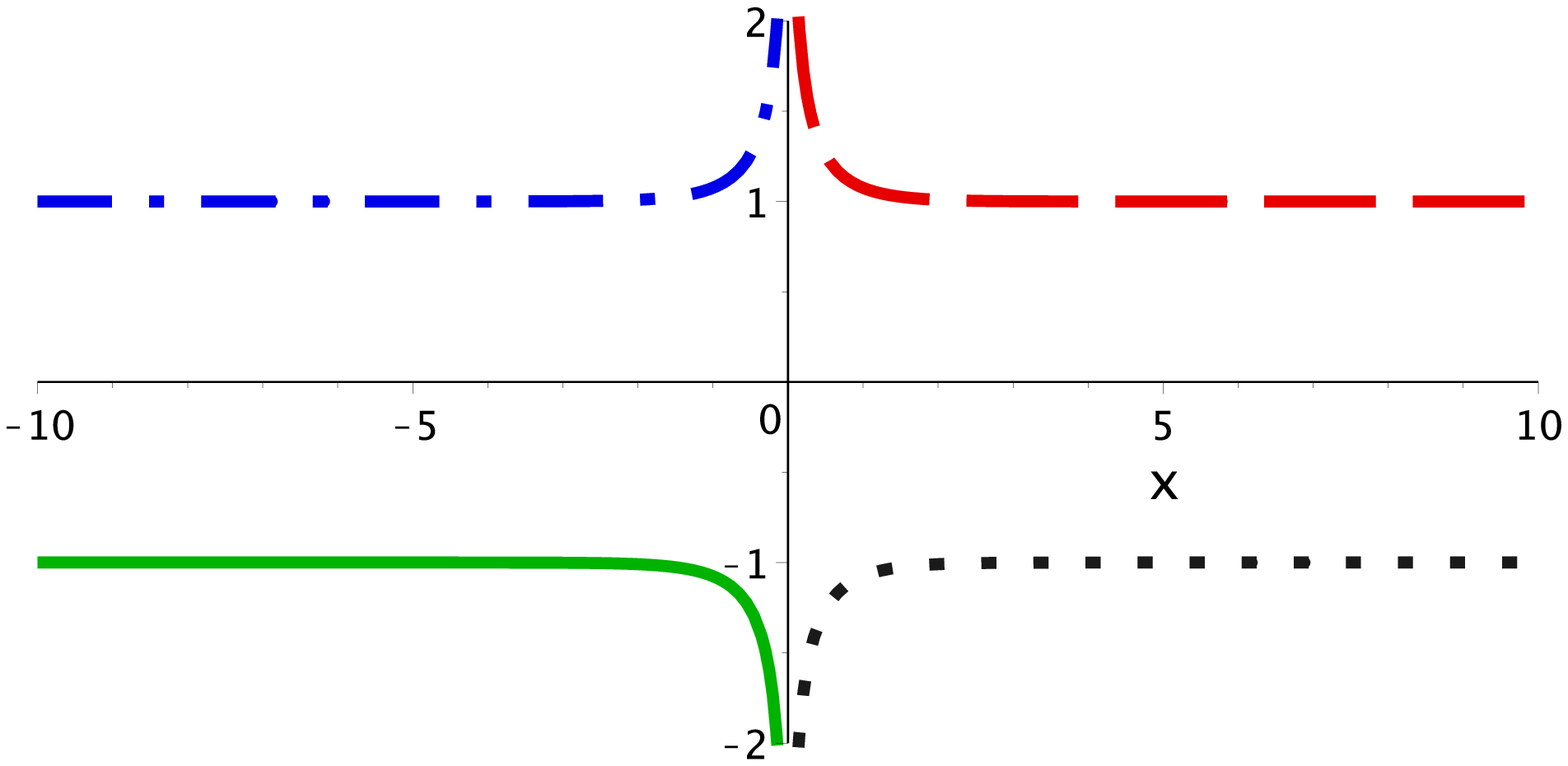}
\caption{The $\phi^{6}$ model on the full line:  a) (left) Static BPS kinks and antikinks on the full line, centered
at $x=0$. Plots are for $\Phi_{1+}(x)$ (red dashed), $\Phi_{2+}(x)$
(blue dash dotted), $\Phi_{1-}(x)$ (black dotted), $\Phi_{2-}(x)$
(green solid line). b) (right) Irregular solutions $\Phi_{3+}(x)$
(red dashed), $\Phi_{4+}(x)$ (blue dash dotted), $\Phi_{3-}(x)$
(black dotted), $\Phi_{4-}(x)$ (green solid line). }
\label{full-line}
\end{figure}
The antikink and kink connecting the minima $\{0,-1\}$ and $\{-1,0\}$
are given, respectively, by $\Phi_{1-}(x)=-\Phi_{1+}(x)$ and $\Phi_{2-}(x)=-\Phi_{2+}(x)$.
Each of these solutions has energy $E=1/4$, and their profiles can
be seen in Fig. \ref{full-line}a. In addition we have four other
solutions of the second-order equation of motion, divergent at $x=0$:
\begin{eqnarray}
\Phi_{3\pm}(x) & = & \pm\sqrt{\frac{1+\coth(x)}{2}}\\
\Phi_{4\pm}(x) & = & \pm\sqrt{\frac{1-\coth(x)}{2}}.
\end{eqnarray}
The profile of these solutions are presented in Fig. \ref{full-line}b.

Stability analysis for $\Phi_{1\pm}$ leads to the following potential
of perturbations:
\begin{eqnarray}
U_{1}(x)=\frac{5}{2}+\frac{3}{2}\tanh(x)-\frac{15}{4}\sech^{2}(x).
\end{eqnarray}
Similarly, for  $\Phi_{2\pm}$ we have
\begin{eqnarray}
U_{2}(x)=\frac{5}{2}-\frac{3}{2}\tanh(x)-\frac{15}{4}\sech^{2}(x). \label{U2}
\end{eqnarray}
\begin{figure}
\includegraphics[width=8cm, height=4cm]{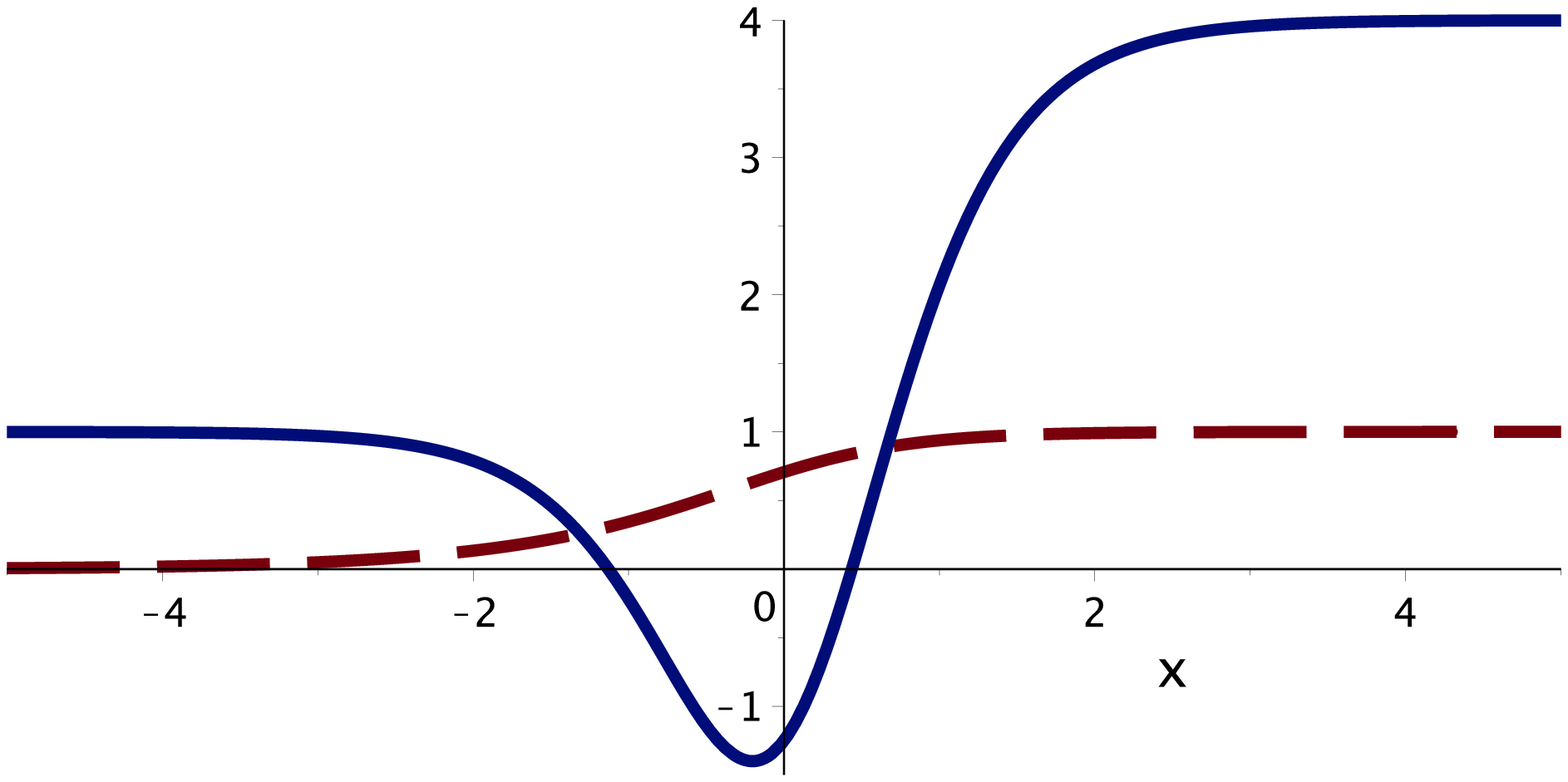} 
\includegraphics[width=8cm, height=4cm]{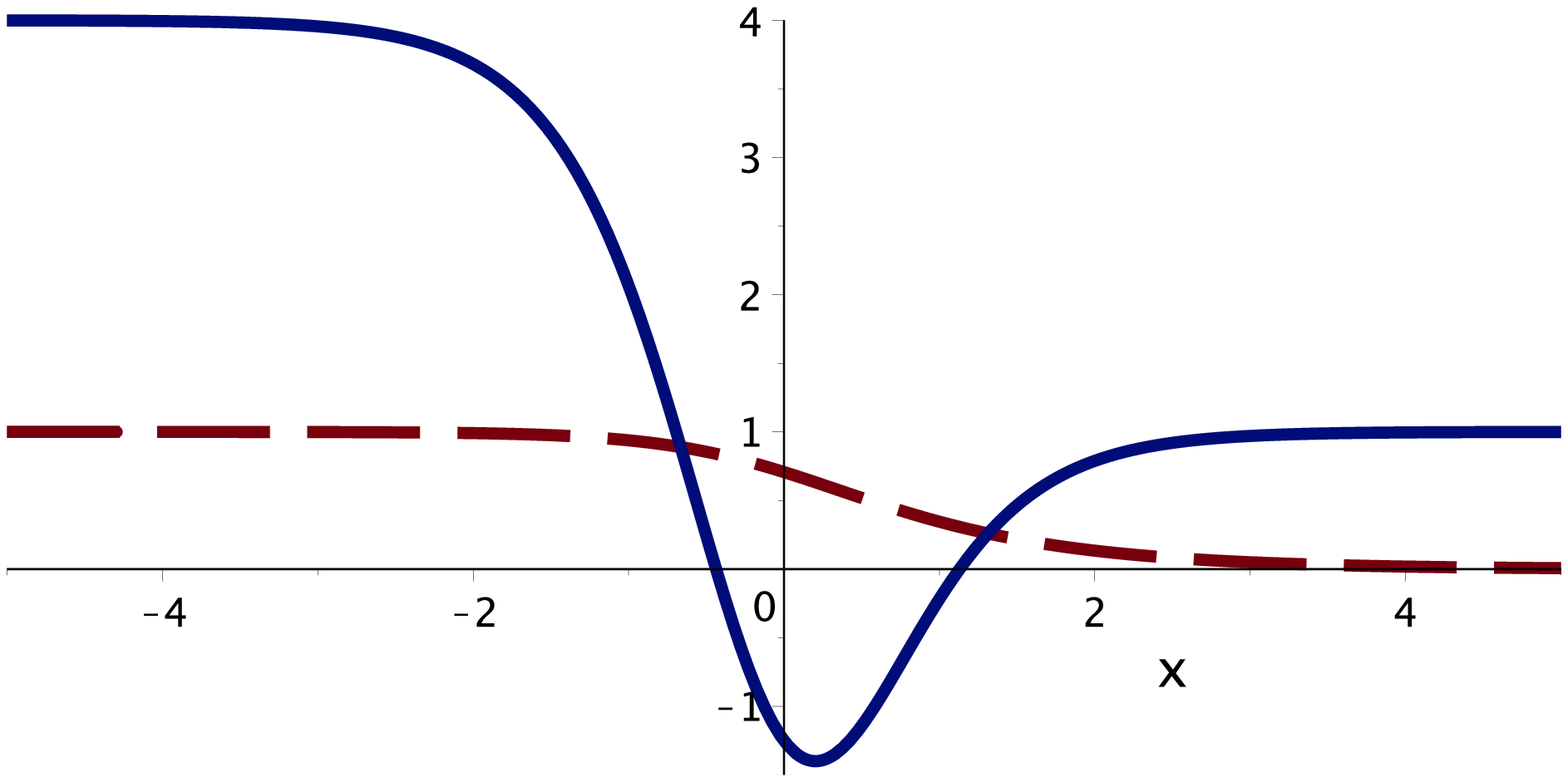}
\caption{The $\phi^{6}$ model on the full line: Schr\"odinger-like potential (solid line) and scalar field profile (traced
line) for a) (left) kink $\Phi_{1+}$ and b) (right) antikink $\Phi_{2+}$.
See Ref. \cite{kk_nonintegr5} }
\label{phi6-pot}
\end{figure}

Figs. \ref{phi6-pot} a-b show the plots of the potentials. There
one can see three characteristic regions: i) $0<\omega^{2}<1$ where
one can look for bound states, ii) $1<\omega^{2}<4$ where there is
a continuum of reflecting states and iii) $\omega^{2}>4$ where \label{reflecting}
there is a continuum of free states. In particular, the analysis of
these states for $\Phi_{1\pm}$ was presented in detail in the Ref. \cite{lohe}
(see also Ref. \cite{mf}). The 
potentials have just one bound state, the zero-mode connected with the
translational symmetry of the solutions. According to the CSW mechanism \cite{kk_nonintegr2}, the absence of a vibrational state would result
in the absence of two-bounce windows in collisions involving kinks
(or antikinks). However, an interesting counterexample was presented
in Ref. \cite{kk_nonintegr5}. In their paper the authors showed the presence
of bounce windows in antikink-kink collisions. 
\begin{figure}
\includegraphics[width=8cm, height=4cm]{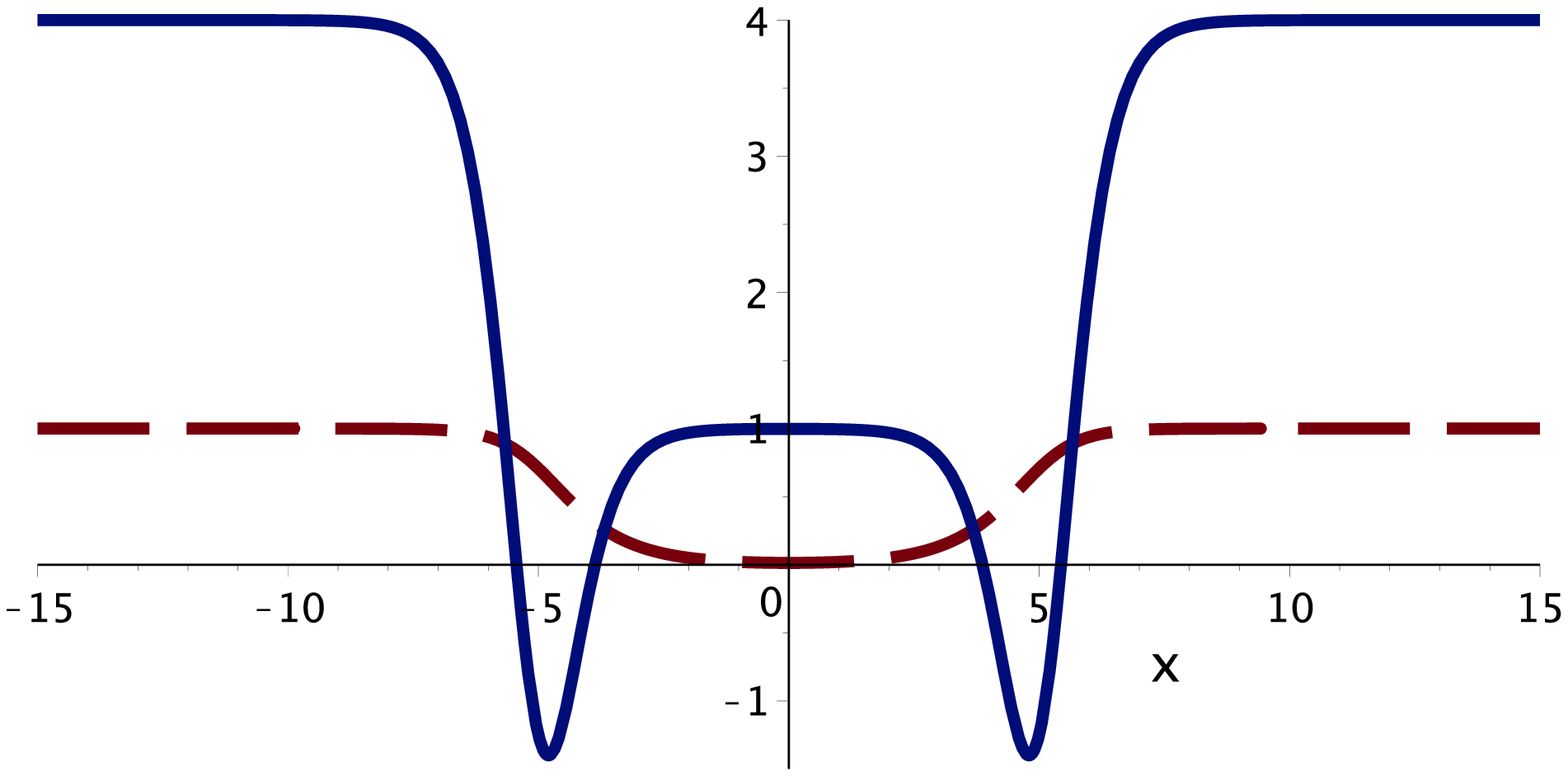} 
\includegraphics[width=8cm, height=4cm]{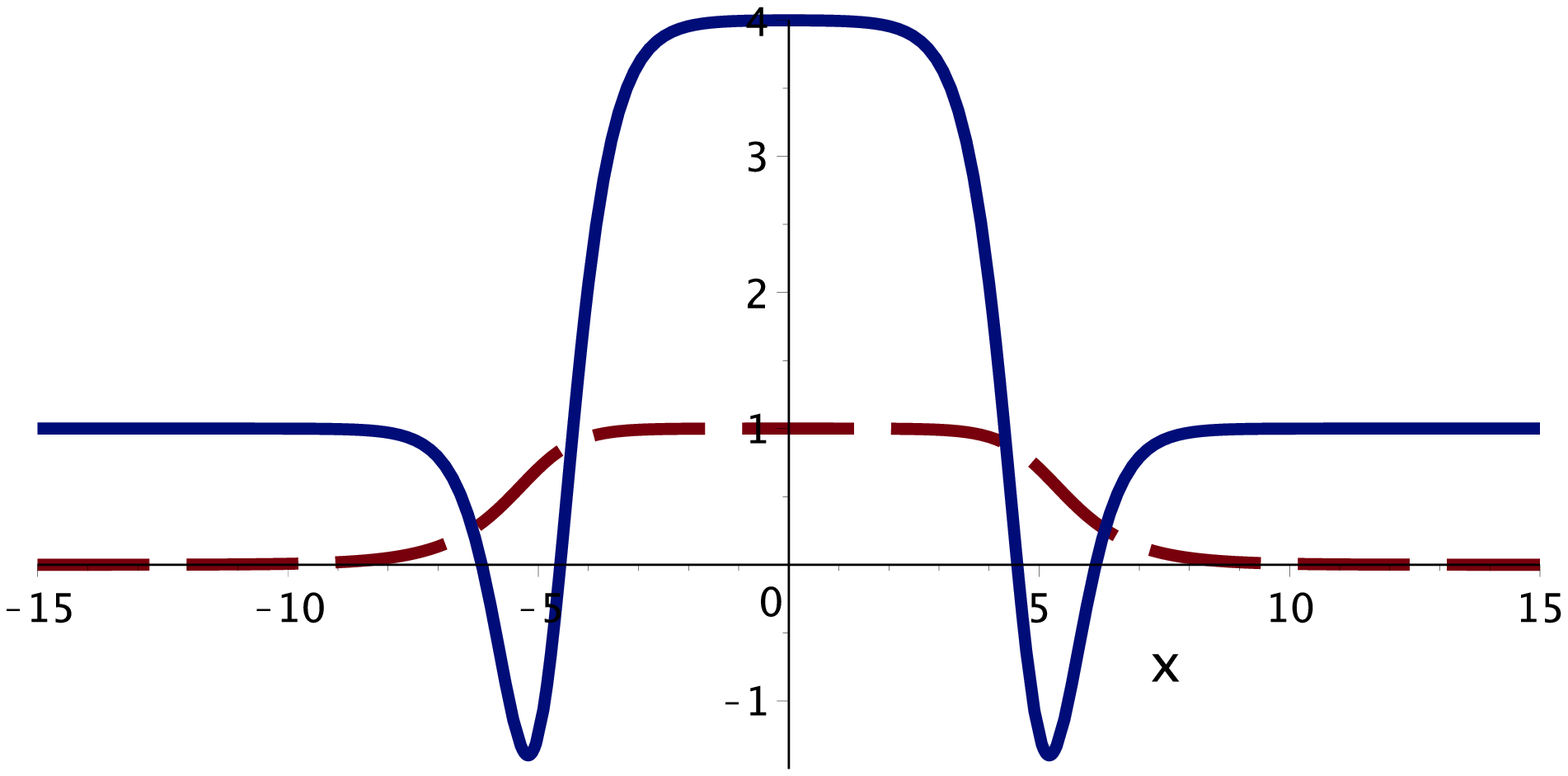}
\caption{The $\phi^{6}$ model on the full line: Schr\"odinger-like potential (solid line) and scalar field profile (traced
line) for a) (left) antikink-kink (composite system $\Phi_{1+}(x-a)+\Phi_{2+}(x+a)$)
and b) (right) kink-antikink (composite system $\Phi_{1+}(x-a)+\Phi_{2+}(x+a)-1$). Here we considered $a=5$.}
\label{potKK}
\end{figure}
The explanation was
that the energy after the initial impact was trapped in the composited
antikink-kink configuration. For example, the composite antikink-kink
system given by $\phi(x)=\Phi_{2+}(x+a)+\Phi_{1+}(x-a)$ has a potential
of perturbations represented in Fig. \ref{potKK}a. This potential
has a tower of bound states. On the contrary, the kink-antikink system
given by $\phi(x)=\Phi_{1+}(x+a)+\Phi_{2+}(x-a)-1$ has the Schr\"odinger-like
potential showed in Fig. \ref{potKK}b. There one has only two quasi-zero
modes. That is the reason why bounce windows are absent in kink-antikink collisions.

\section{The $\phi^6$ model on the half-line}
\label{half}

\begin{figure}
	\begin{centering}
\includegraphics[width=16cm]{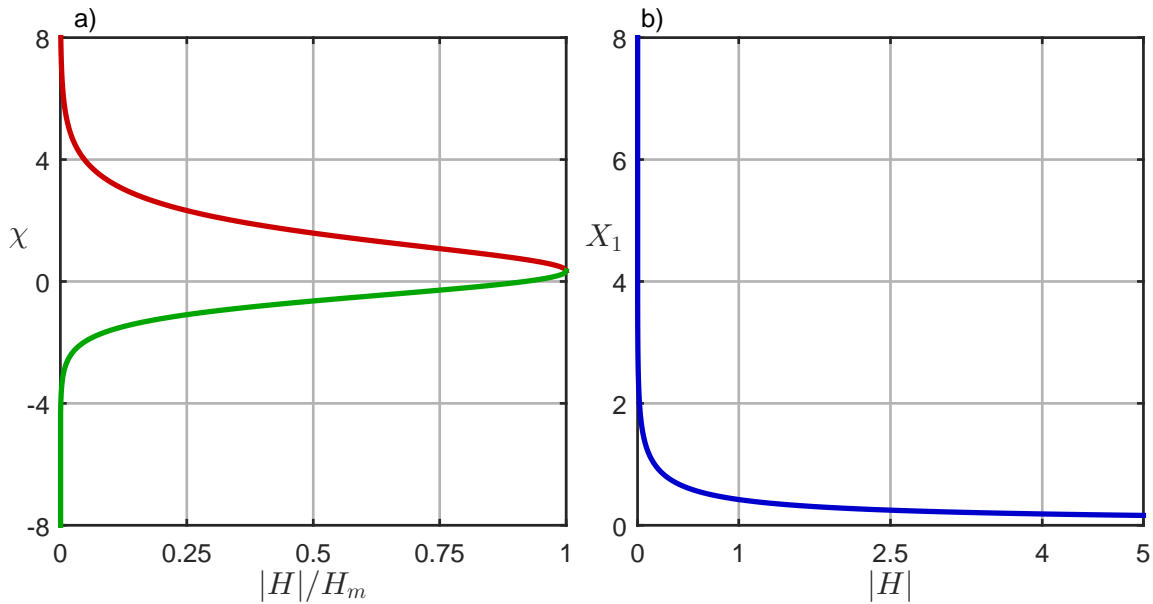}
\caption{ a) The two real solutions of Eq. (\ref{X0b}) for $\chi(H)$: $X_0$ (red) and $\hat{X}_0$ (green); b) the unique real solution  $X_{1}(H)$ of  Eq. (\ref{X1b}). }
\label{X_0versusH}
	\end{centering}
\end{figure}

Now we consider the $\phi^6$ model on the half-line $-\infty<x<0$. Firstly we will consider the solutions  given by
\begin{eqnarray}
\varphi_{1+}(x) &=& \sqrt{\frac{1+\tanh[x-\chi(H)]}{2}},\\
\varphi_{4+}(x) &=& \sqrt{\frac{1-\coth[x-X_1(H)]}{2} },
\end{eqnarray}
which correspond respectively to  $\Phi_{1+}(x)$ and $\Phi_{4+}(x)$ on the full line.
The Neumann condition  (Eq. (\ref{neumann})), impose that
\begin{eqnarray}
\frac{\sech^2\chi}{\sqrt{8-8\tanh\chi}}=H,
\label{X0}\\
\frac{\coth X_1^2 - 1}{\sqrt{8+8\coth X_1}}=H.
\label{X1}
\end{eqnarray}
The Eqs. (\ref{X0}) and (\ref{X1}) have no real solutions for $H<0$. The Eq. (\ref{X0}) has two solutions for each value of $H$ in the interval $0 \le H \le H_m $, where $H_m=2/\sqrt{27}$. We will call these solutions  of  $X_0(H)$ and $\hat{X}_0(H)$ with $\hat{X}_0\le X_{0}$. For $H=H_m$ we have $\hat{X}_0 = X_{0}$.
As a comparison, for the $\phi^4$ model on the half-line we have  \cite{dorey1} $|\hat{X}_0|=|X_{0}|$ and $H_m=1$. In addition the Eq. (\ref{X1})  has a unique real solution for $X_1(H)$ in the interval $0<H<\infty$.   Similar reasoning can be done with the solutions
\begin{eqnarray}
\varphi_{2+}(x) &=& \sqrt{\frac{1-\tanh[x+\chi(H)]}{2}},\\
\varphi_{3+}(x) &=& \sqrt{\frac{1+\coth[x+X_1(H)]}{2} }.
\end{eqnarray}
In this case the Neumann condition  (Eq. (\ref{neumann})) gives also Eqs. (\ref{X0}) and (\ref{X1}).
\begin{eqnarray}
\frac{\sech^2\chi}{\sqrt{8-8\tanh\chi}}=-H,\label{X0b}\\
\frac{\coth X_1^2 - 1}{\sqrt{8+8\coth X_1}}=-H.
\label{X1b}
\end{eqnarray}
The Eqs. (\ref{X0b}) and (\ref{X1b}) have no real solutions for $H>0$. The Eq. (\ref{X0b}) has two solutions, $X_0(H)$ and $\hat{X}_0(H)$ in the interval $-H_m \le H \le 0 $ whereas Eq. (\ref{X1b}) has one solution for each value of $H<0$. The Fig. \ref{X_0versusH}a presents the behavior the parameters $X_0$ and $\hat X_0$  as function of $|H|/H_m$. The Fig. \ref{X_0versusH}b presents the parameter $X_1$  as function of $|H|$. In the figures the signal of $H$ depends on the particular $\varphi_{i+}$, $i=1..4$.

We can classify the solutions depending on the BPS equations they satisfy. Then we have 
\begin{eqnarray}
\phi_{1}(x) & = & \sqrt{\frac{1+\tanh(x-X_{0})}{2}} \label{phi1}\\
\phi_{2}(x) & = & \sqrt{\frac{1+\tanh(x-\hat{X}_0)}{2}} \label{phi2}\\
\phi_{3}(x)  &= & \sqrt{\frac{1+\coth(x+X_{1})}{2}}. \label{phi3}
\end{eqnarray}
which satisfy BPS Eq. (\ref{bps1}). Also we have 
\begin{eqnarray}
\tilde{\phi}_{1}(x) & = & \sqrt{\frac{1-\tanh(x+\hat{X}_0)}{2}} \label{p1til}\\
\tilde{\phi}_{2}(x) & = & \sqrt{\frac{1-\tanh(x+X_{0})}{2}}, \label{p2til}\\
\tilde{\phi}_{3}(x) & = & \sqrt{\frac{1-\coth(x-X_{1})}{2}},
\label{p3til}
\end{eqnarray}
which satisfy BPS Eq. (\ref{bps2}). Note that $\phi_1,\phi_2$ and $\tilde\phi_3$ are solutions for $H>0$ whereas  $\tilde\phi_1,\tilde\phi_2$ and $\phi_3$ are solutions for $H<0$. 
\begin{figure}
	\includegraphics[width=16cm]{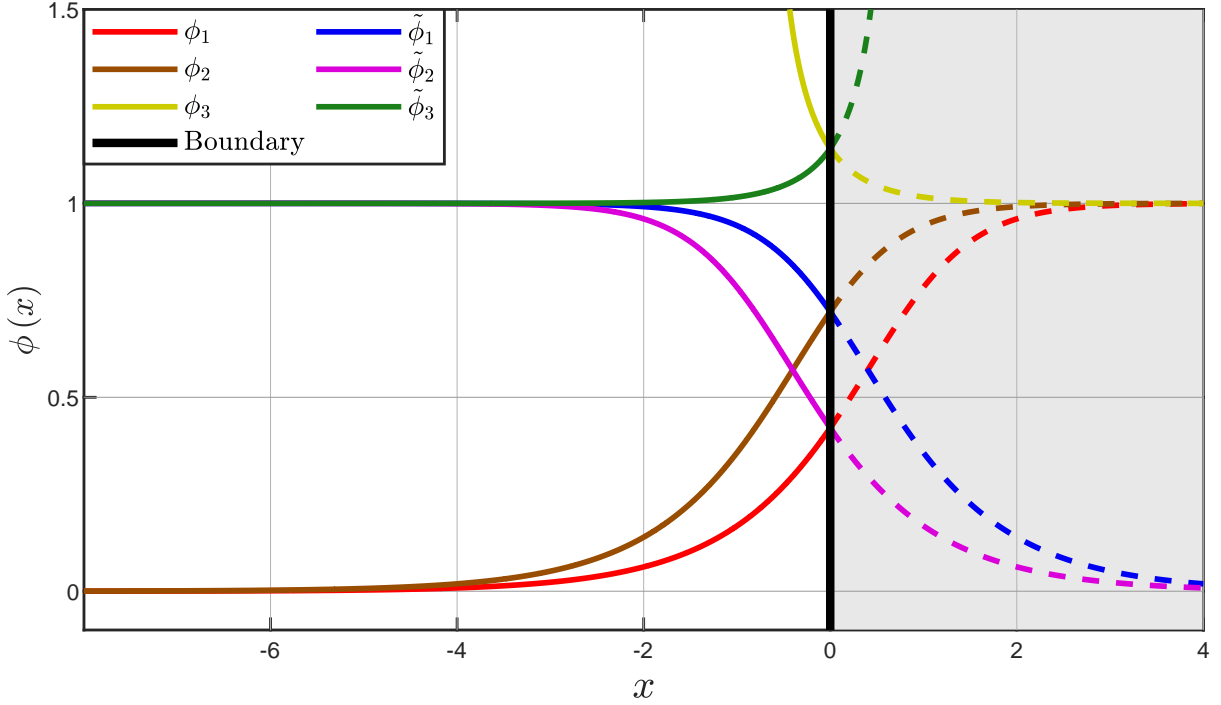}
	\caption{ Static solutions for $|H|=0.90H_m$.  }
	\label{perfil-phi6}
\end{figure}

The profiles of the solutions in the region $x<0$ are depicted in the Fig. \ref{perfil-phi6}. Solution $\phi_{3}(x)$ is not regular for $x<0$ and will not be considered here.  The energy of static solutions can be found using Eq. (\ref{Ephi}). Then, the energies for solutions (\ref{phi1} - \ref{phi2}) are given by 
\begin{eqnarray}
\label{Ephia}
E[\phi_i(x)]=\frac{3}{4}\phi_{i}(0)^4 - \frac{1}{2}\phi_{i}(0)^2,
\end{eqnarray}
with $i=1\dots2$ whereas those for solutions (\ref{p1til} - \ref{p3til}) are given by
\begin{eqnarray}
\label{Ephib}
E[\tilde{\phi}_i(x)]=\frac{1}{4}+ \frac{1}{2}\tilde{\phi}_{i}(0)^2-\frac{3}{4}\tilde{\phi}_{i}(0)^4
\end{eqnarray}
with $i=1\dots3$. The Eqs. (\ref{Ephia}) and (\ref{Ephib}) shows that the energies presents an implicit  dependence with $H$ through of $ \phi_i(0) $ and $ \tilde{\phi}_i(0) $.

\begin{figure}[h]
	\begin{centering}
		\includegraphics{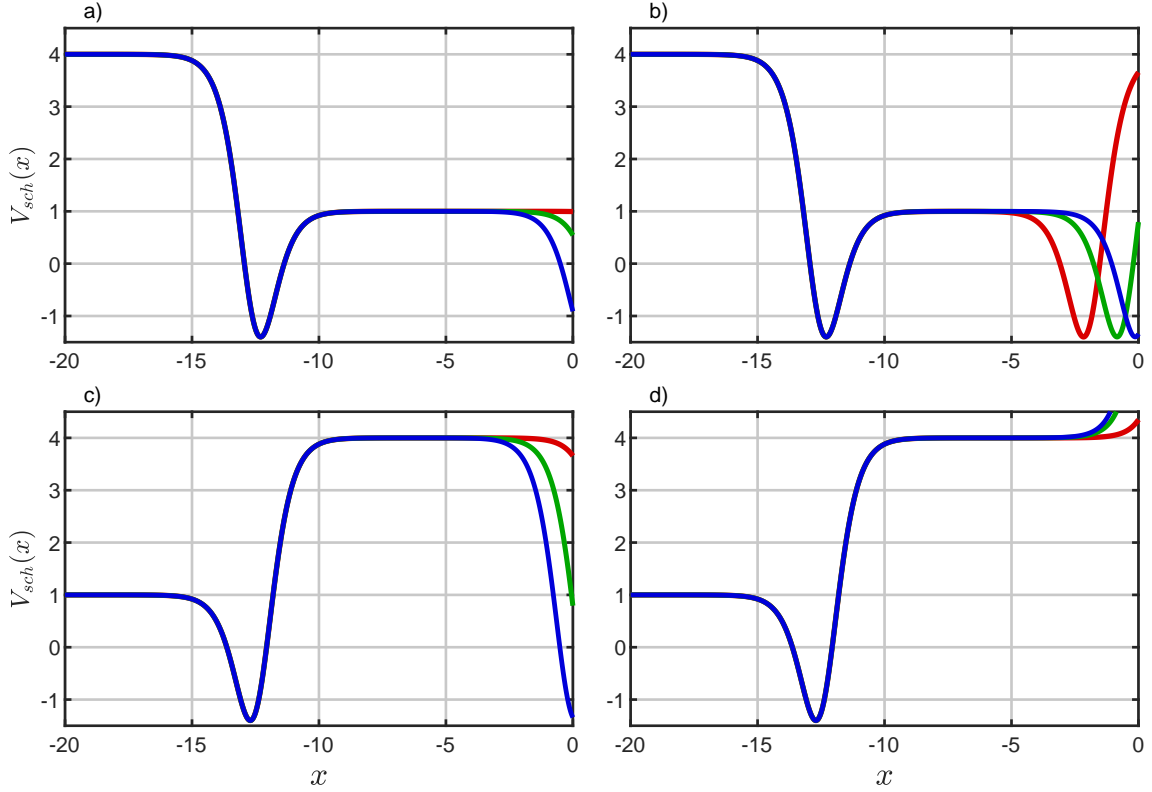} 
		\par\end{centering}
	\caption{ Schr\"odinger-like potential  for static configurations: a) $\Phi_{2+}(x-a)+\phi_1(x)$ b) $\Phi_{2+}(x-a)+\phi_2(x)$ c) $\Phi_{1+}(x-a)+\tilde{\phi}_1(x)-1$ d) $\Phi_{1+}(x-a)+\tilde{\phi}_2(x)-1$. The lines are for $|H|=0.05H_m$ (red), $|H|=0.50H_m$ (green) and $|H|=0.95H_m$ (blue). We fix $a=-12.50$. }
	\label{Vsch_bound}
\end{figure}

The Fig. \ref{Vsch_bound} shows the Schr\"odinger-like potential  for some static configurations of antikink-boundary and kink-boundary for several values of $H$.  Note from the figure that potential is composed of two parts: one, with a hole centered at $x_0$ (chosen as $-12.5$ in the figure), characterizes the kink/antikink defect (known to have only the translational mode $\omega^2=0$); the second one, close to $x=0$, represents the influence of the boundary. Note also that for $\omega^2>4$ one has a continuum of states. The possibility of vibrational states depends on the boundary. The Figs.  \ref{Vsch_bound}a-b shows that, for the boundaries $\phi_1$ and $\phi_2$, vibrational states must be searched for $\omega^2\le 4$. On the other hand, the potentials of the Figs.  \ref{Vsch_bound}c-d show that true vibrational states can occur for $\omega^2< 4$; for  $1<\omega^2< 4$ we have mestastable states, similar to the reflecting states in the full line of Sect. \ref{reflecting}.

\begin{figure}[h]
	\begin{centering}
		\includegraphics[width=15cm]{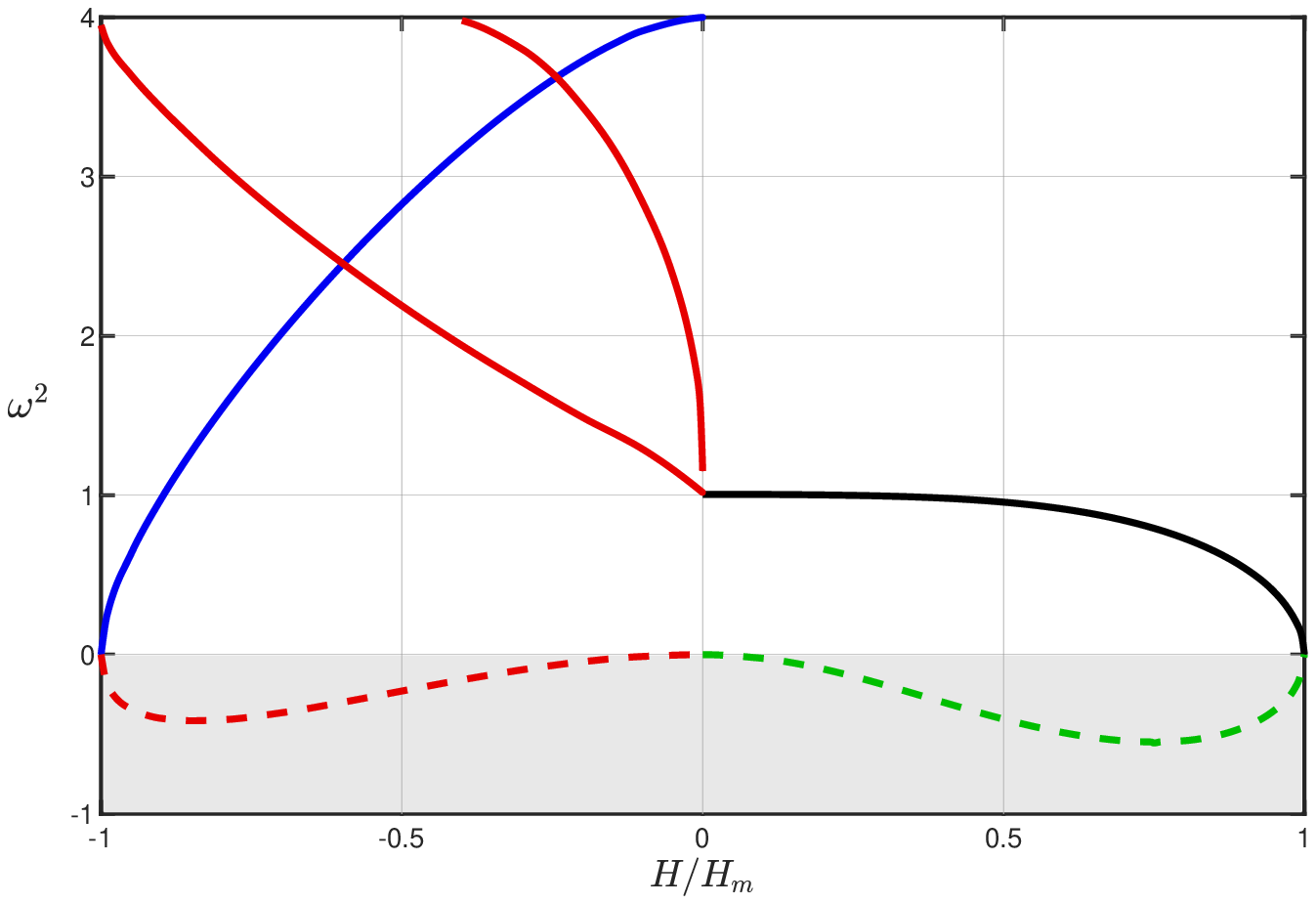} 
	\par\end{centering}
	\caption{The squared frequencies $\omega^2$ of localised boundary modes as a function of H. The curves are for solutions $\phi_1(x)$ (black solid line) and $\phi_2(x)$ (green dashed line), $\tilde{\phi}_1(x)$ (blue solid line) and $\tilde{\phi}_2(x)$ (red dashed and solid lines).}
	\label{states}
\end{figure}

In order to study the occurrence of localized boundary modes, we solve numerically the Schr\"odinger-like equation when the defect is too far from the boundary. The Fig. \ref{states} shows the squared frequency of these modes  as a function of $H$. For 
$0<H<H_m$, we analyzed the solutions $\phi_1(x)$, $\phi_2(x)$ and $\tilde{\phi}_3(x)$. For $\phi_1(x)$ we have one vibrational state with $0<\omega^2<1$ and the solution is stable. For $\phi_2(x)$ we have one mode with $\omega^2<0$, meaning instability.  The solution $\tilde{\phi}_3(x)$ has no bound states, meaning also no signal of instability.
For $-H_m<H<0$ we analyzed the solutions $\tilde{\phi}_1(x)$ and $\tilde{\phi}_2(x)$. For $\tilde{\phi}_1(x)$ there is one state with $\omega^2>0$ and the solution is stable. The classification of this state depends on the value of $H$. The Fig. \ref{states} shows that for $-0.90\lesssim H/H_m<0$ we have $1<\omega^2<4$. When we consider the whole system antikink-boundary, this is in fact a metastable state, as already discussed.  For $-1<H/H_m\lesssim -0.90$ we have  $0<\omega^2<1$ and the solution is a true vibrational state of the boundary. The solution $\tilde{\phi}_2(x)$ has three bound states, one of them with $\omega^2<0$, meaning instability. 

\begin{figure}
\includegraphics[width=15cm]{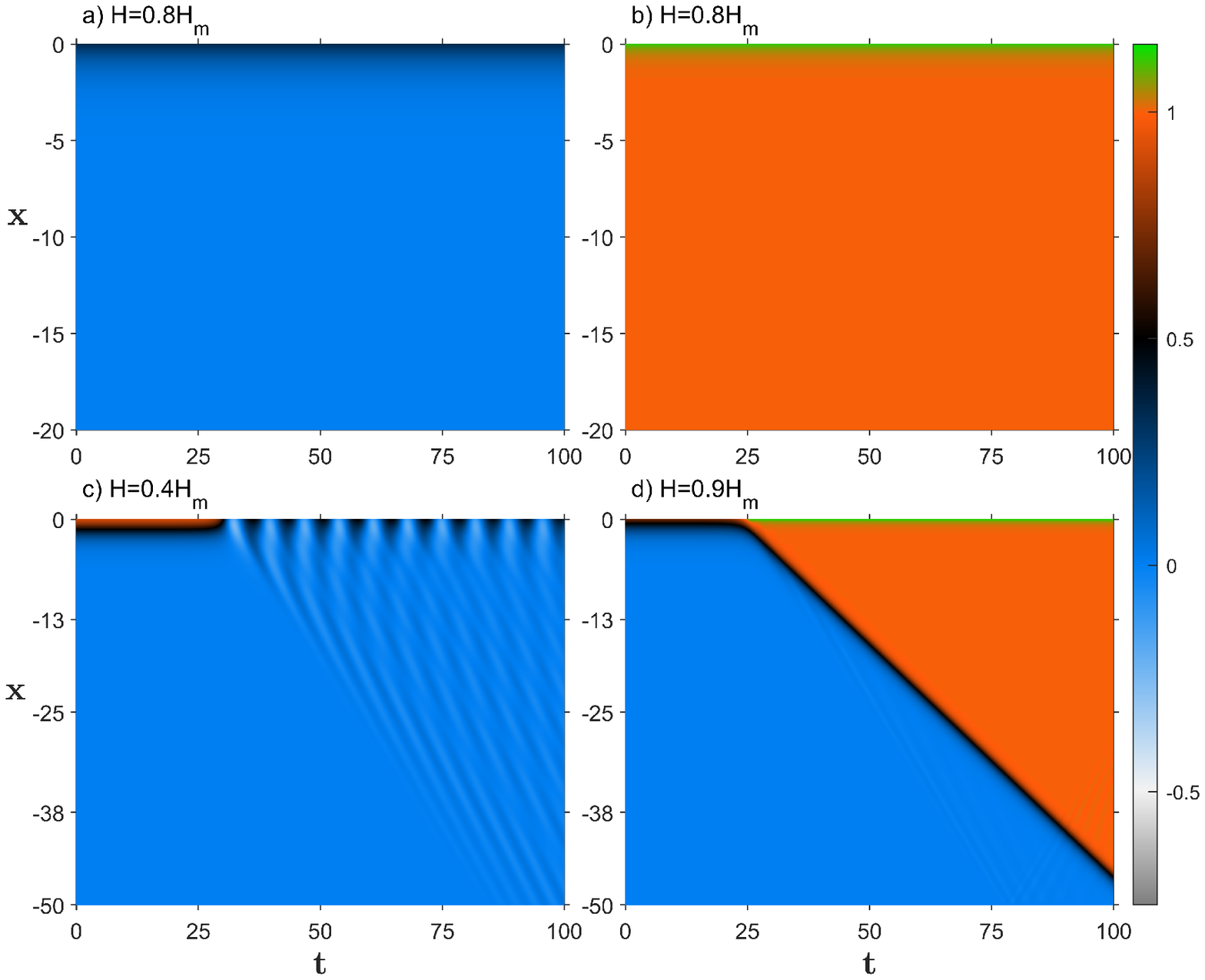}
\caption{Stability analysis of the boundaries: a) unchanging of $\phi_1(x)$ configuration, compatible with stability; b) unchanging of $\tilde\phi_3(x)$ configuration, compatible with stability; c) emission of radiation by the unstable $\phi_2$ boundary; d) emission of a kink by the unstable $\phi_2$ boundary. The values of $H$ are shown in the figures.}
\label{bound}
\end{figure}

The stability of the boundaries was also analyzed numerically, considering only each boundary isolated, without any scattering process. Set $\xi(x)$ one of the boundaries $\phi_1(x), \tilde{\phi}_1(x)$, $\phi_2$ and $\tilde{\phi}_3(x)$. The initial conditions are
\begin{eqnarray}
\phi(x,0)&=&\xi(x),\\
\dot\phi(x,0)&=& 0.
\end{eqnarray}
For the numerical solutions we used a $4^{th}$-order finite-difference
method on a grid $N=2048$ nodes with the ``infinity" at $x_{min}=-100$ and a spatial step
$\delta x\approx 0.05$. For the time dependence we used a $6^{th}$ order sympletic integrator
method, with a time step $\delta t=0.02$. We found no modification of the boundaries  $\phi_1(x), \tilde{\phi}_1(x)$ and $\tilde{\phi}_3(x)$ with the increasing of $H$ even for large times, which is compatible with stability (see the Figs. \ref{bound}a-b for the boundaries $\phi_1(x)$ and $\tilde{\phi}_3(x)$). On the other hand the instability of the ${\phi}_2(x)$ shows up in the simulations. One sees, for $H\lesssim0.72$, the emission of radiation by the boundary, as shown for instance in the Fig. \ref{bound}c. For $H\gtrsim0.72 H_m$ one has the production of a kink by the boundary, as shown for instance in the Fig. \ref{bound}d. For such high values of $H$ there is no emission of radiation before or after the production of the kink. In this case the effect of instability shows up abruptly, with no signal of instability thereafter. This is compatible with a decaying of the 
unstable $\phi_2$ boundary to the stable $\phi_1$. 

In the remaining of this work we will study of boundary scattering in the $\phi^6$ model. For this purpose only the regular and stable solutions $\phi_1(x), \tilde{\phi}_1(x)$ and $\tilde{\phi}_3(x)$ will be considered. We will see that even these stable boundaries can emit radiation, produce a kink/antikink or even change their nature due to the scattering. Clearly this is a consequence of the interaction defect-boundary, not a signal of instability.

\section{Antikink-boundary scattering}

Here we will consider scattering with an antikink in the sector $(0,1)$ with the boundary. The Fig. \ref{perfil-phi6} shows that an antikink in this sector can be connected with solutions  $\phi_1(x)$ and $\phi_2(x)$, but only $\phi_1(x)$ is regular and stable. This means to restrict the parameter $H$ to the range $0<H<H_{m}$. The antikink is the static solution $\Phi_{2+}(x)$ given by the Eq. (\ref{Phi2+}), now boosted with a velocity $v_i$ with $x(t=0)=a$ and given by:
\be
\Phi_{2+}(v_i,x,t)=\sqrt{\frac{1-\tanh(\gamma(x-v_it-a))}{2}}
\ee
with $\gamma=1/\sqrt{1-v_i^2}$. The initial conditions are
\begin{eqnarray}
\phi(x,0)&=&\Phi_{2+}(v_i,x,0)+\phi_{1}(x),\\
\dot\phi(x,0)&=& \dot\Phi_{2+}(v_i,x,t)|_{t=0}.
\end{eqnarray}
For the numerical solutions we used $a=-12.5$ and a $4^{th}$-order finite-difference
method. For the time dependence we used a $6^{th}$ order sympletic integrator
method.

\begin{figure}
\begin{centering}
\includegraphics[scale=0.8]{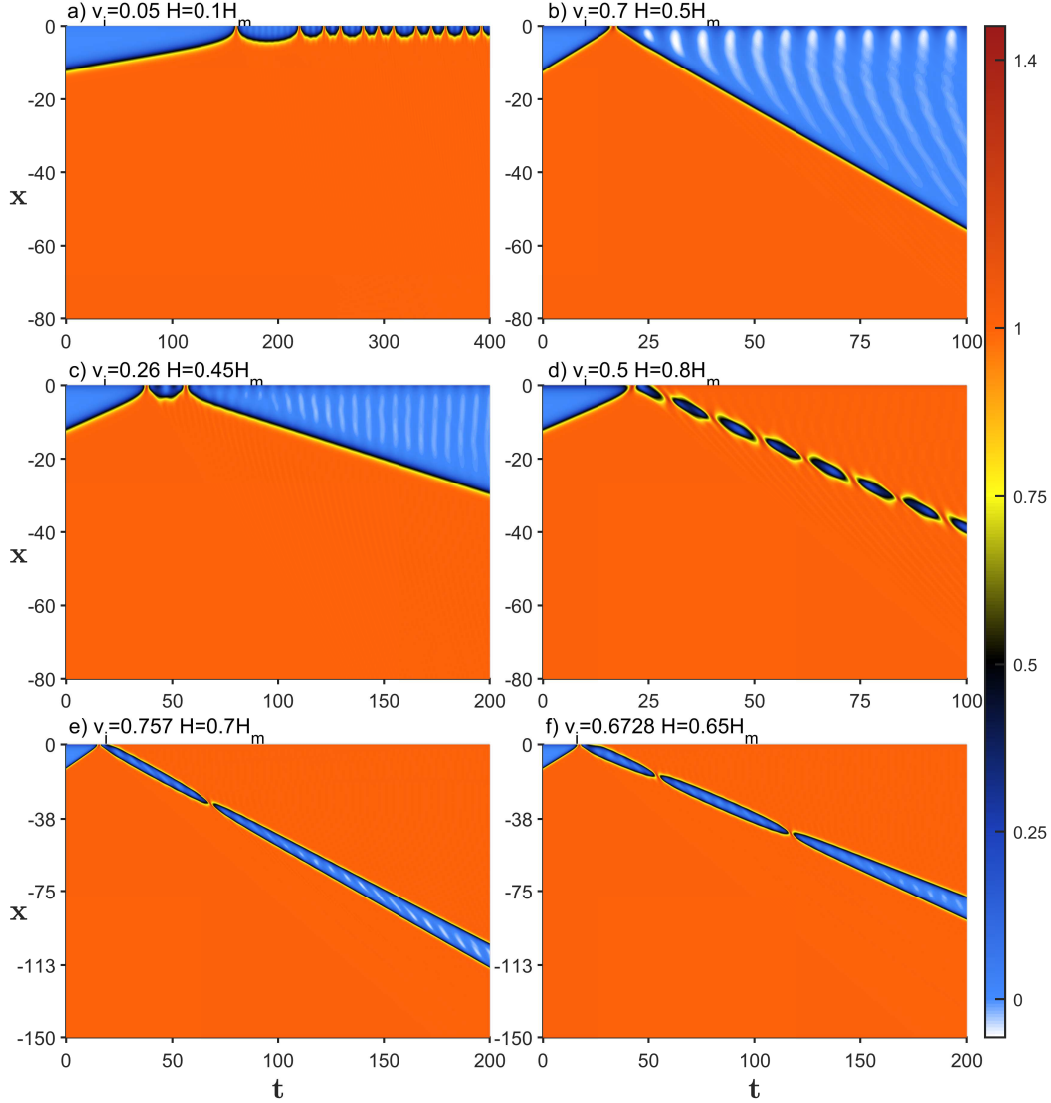}
\par\end{centering}
\caption{$\Phi_{2+}(x)$ antikink scattering with $\phi_1$ boundary, showing a) bion state at the boundary, b) inelastic scattering  (one-bounce) with the boundary, c) two-bounce collision with the boundary, d) production of a $\Phi_{1+}(x)$ kink from the boundary, forming a composed oscillating state, e) production of a $\Phi_{1+}(x)$ kink from the boundary, showing the antikink colliding once with the formed kink,  f) production of a $\Phi_{1+}(x)$ kink from the boundary, showing the antikink colliding twice with the formed kink. One can localize the structure of the output field profile at the boundary in the next Fig. \ref{mosaico}.  
\label{boundary}}
\end{figure}
The Fig. \ref{boundary} shows the richness of possibilities of the scattering products. The realization of each scenario depends on the parameters $(v_i,H)$. The Fig. \ref{boundary}a shows a bion state at the boundary. Fig. \ref{boundary}b shows an inelastic scattering  (one-bounce) with the boundary. Fig. \ref{boundary}c shows a two-bounce collision with the boundary.  The Fig. \ref{boundary}d shows the production of a $\Phi_{1+}(x)$ kink by the boundary together with an inelastic scattering of the antikink. Note that the antikink-kink pair do not separate completely, forming a composed oscillating state. The Figs. \ref{boundary}e  (\ref{boundary}f) show cases where the produced $\Phi_{1+}(x)$ kink scatter once (twice) with the reflected antikink.   After this collision, the separation between antikink and kink grows slowly with time, favoring the exchanging of radiation between the antikink-kink pair. 

 The contribution of the bound states to bion and two-bounce windows can be seen in the Figs.  \ref{boundary}a and \ref{boundary}c.  There we can see that between bounces there is emission of radiation whose frequency decreases with $H$. This agrees with the Fig.  \ref{states} (black curve). The same effect is observed  in the pattern of the oscillations at the boundary after the scattering (compare the Figs. \ref{boundary}c and \ref{boundary}b).

\begin{figure}
	\includegraphics[scale=0.9]{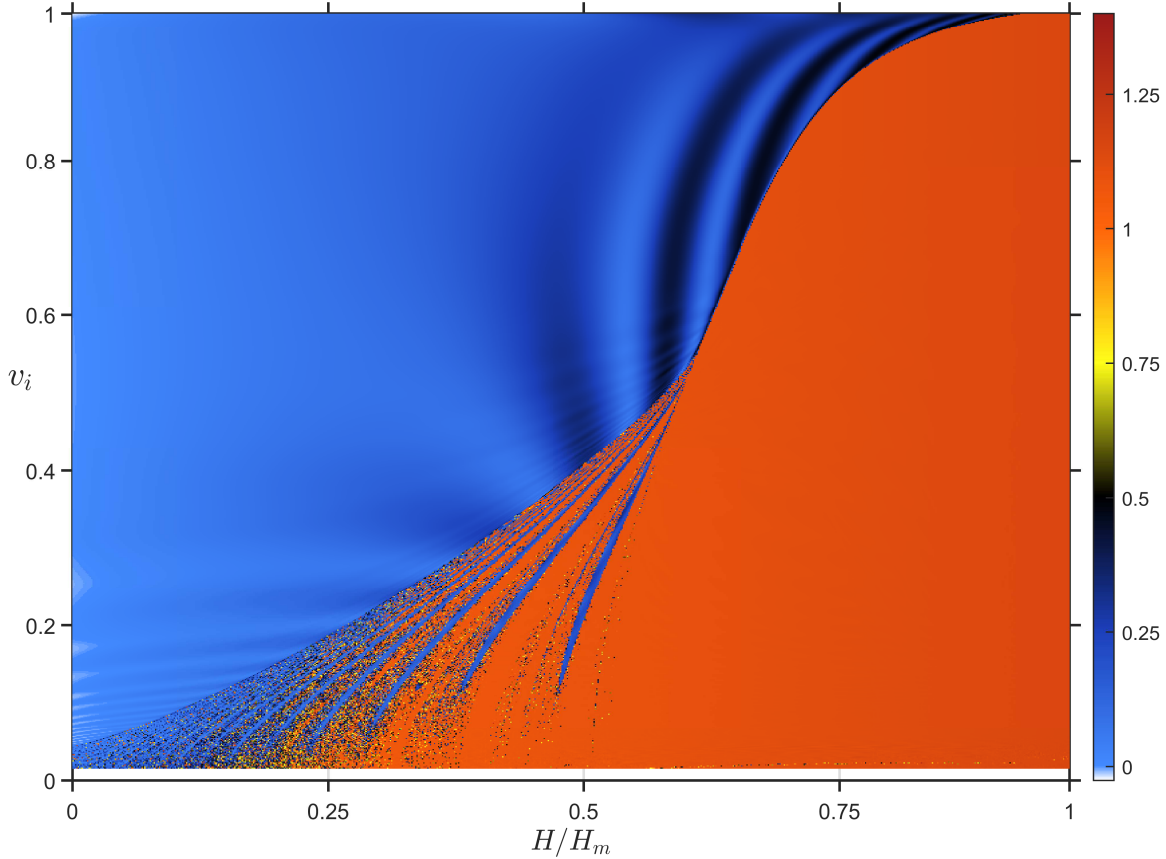}
    \caption{$\Phi_{2+}(x)$ antikink scattering with $\phi_1$ boundary: final state of the scalar field $\phi\left( 0,t_{f}\right)$, with $t_{f}={|a|}/{v_{i}}+100$.  }
	\label{mosaico}
\end{figure}

\begin{figure}
\begin{centering}
\includegraphics[width=10cm]{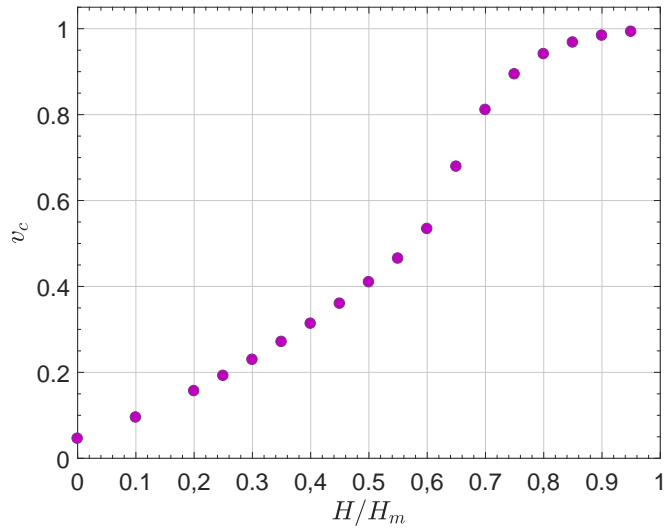}
\par\end{centering}
\caption{$\Phi_{2+}(x)$ antikink scattering with $\phi_1$ boundary: critical velocity $v_c$ as a function of $H$. For $v_i>v_c$ there is no possibility of formation of a travelling kink at the boundary (see also Fig. \ref{mosaico}).}
\label{vcversusH}
\end{figure}

The structure of our results on the boundary is presented in the Fig. \ref{mosaico}.
There one can see, in the bidimensional $(H, v_i)$ phase space, the final state of the scalar field at $x=0$.  In the blue region the antikink is inelastically scattered by the boundary after one bounce. The red region characterizes the production of a $\Phi_{1+}(x)$ kink by the boundary.  In the figure we can observe the dependence with $H$ of the structure of bounce windows. There the bounce windows are characterized by the finite intervals in blue ($\phi(0,t_f) \sim 0$) in the region $0<H\lesssim 0.6 H_m$, $v_i \lesssim 0.55$.  The frontier red-blue characterize the critical velocity $v_c$, above which the antikink collides only once with the boundary, thereafter retreating to 
$x\to-\infty$ (one can better visualize this in the Fig. \ref{vcversusH}).  This is the analogous situation of a one-bounce solution for antikink-kink scattering in the full line.  For velocities $v_i<v_c$ there are several possibilities: i) bion states with bounce windows for $v_i\lesssim v_c$. This is mostly observed for small values of $H$ (left side of Fig. \ref{mosaico}). ii) production of kink by the boundary for large values of $H$ (right side of Fig. \ref{mosaico}).

Fig. \ref{vcversusH} shows the behavior
of the critical velocity as a function of the parameter $H$. Note that
the larger is $H$, the larger is the critical velocity. In particular, in the limit
$H\rightarrow H_{m}$ we have $v_{c}\rightarrow1$ and the formation of a travelling kink at the boundary. 
Note also that the curve $v_{c}$ {\it versus} $H$ presents an inflection point
around $H \simeq 0.6H_{m}$, which coincides with the maximum value where two bounce windows are observed. For $H\gtrsim0.6H_{m}$, $v_c$ grows with $H$ at a higher rate, favoring the occurrence of kink emission by the boundary.  

\begin{figure}[H]
\begin{singlespace}
\end{singlespace}
\begin{centering}
\includegraphics{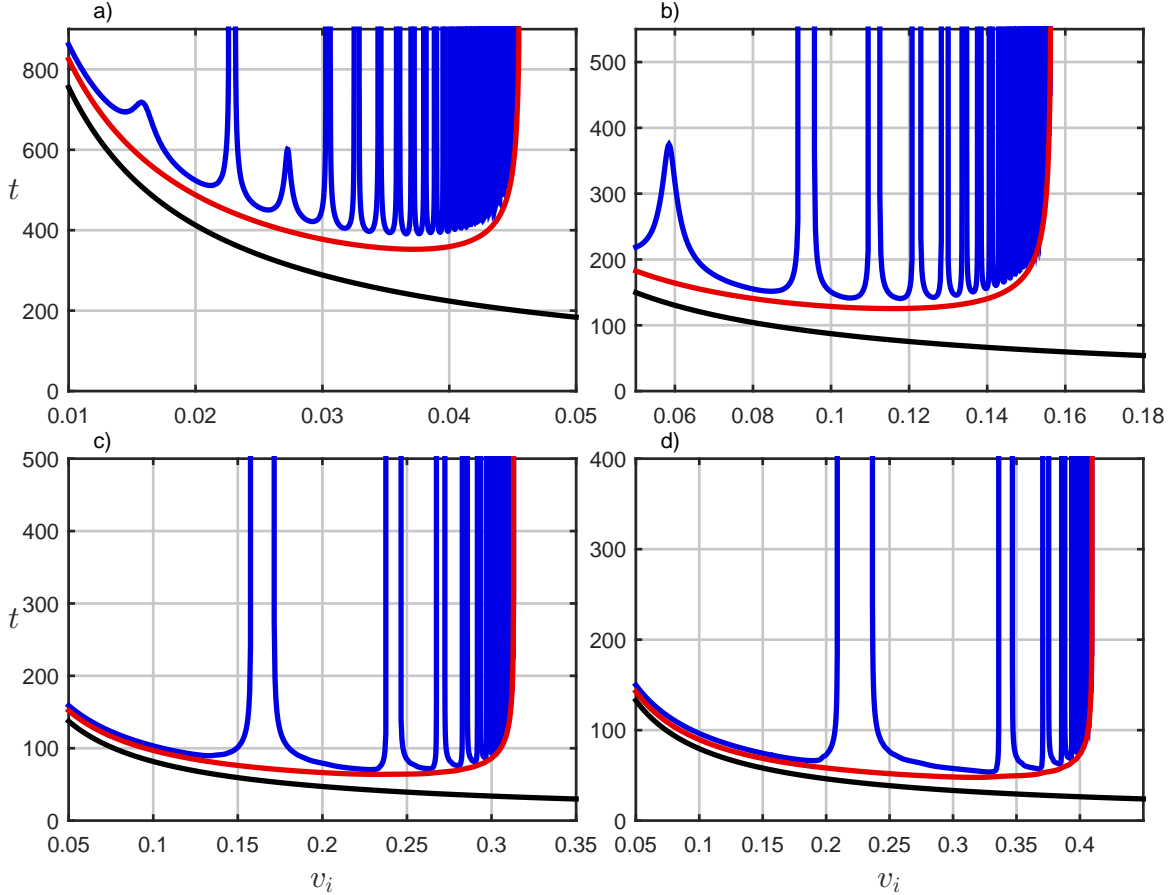}
\end{centering}
\caption{$\Phi_{2+}(x)$ antikink scattering with $\phi_1$ boundary: time of the first (black), second (red) and third (blue) bounce {\it versus} initial velocity for (a) $H=0$, (b) $H=0.2H_{m}$,
(c) $H=0.4H_{m}$, and (d) $H=0.5H_{m}$. }
\label{vnb}
\end{figure}

The effect of $H$ on the structure of two-bounce windows is described
in the Figs. \ref{vnb}a-\ref{vnb}d, where we see the times for the first three bounces as a function of the initial velocity. The two-bounce windows are delimited by divergences in the curves of the time of the third bounce. First of all, we observe that
when $H=0$ we recover the structure of two-bounce windows presented
in antikink-kink scattering for the $\phi^{6}$ model in the full
line, as described in the Ref. \cite{kk_nonintegr5}. As $H$ grows from zero,
also grows the critical velocity $v_{c}$ that separated one-bounce
inelastic scattering from bion states. We note also that with the
grow of $H$ the two-bounce windows accumulate around
$v_i=v_{c}$. For $H\simeq 0.6H_{m}$ they disappear, remaining
just the phenomena of kink production on the boundary.


\begin{figure}
\includegraphics[width=5.5cm]{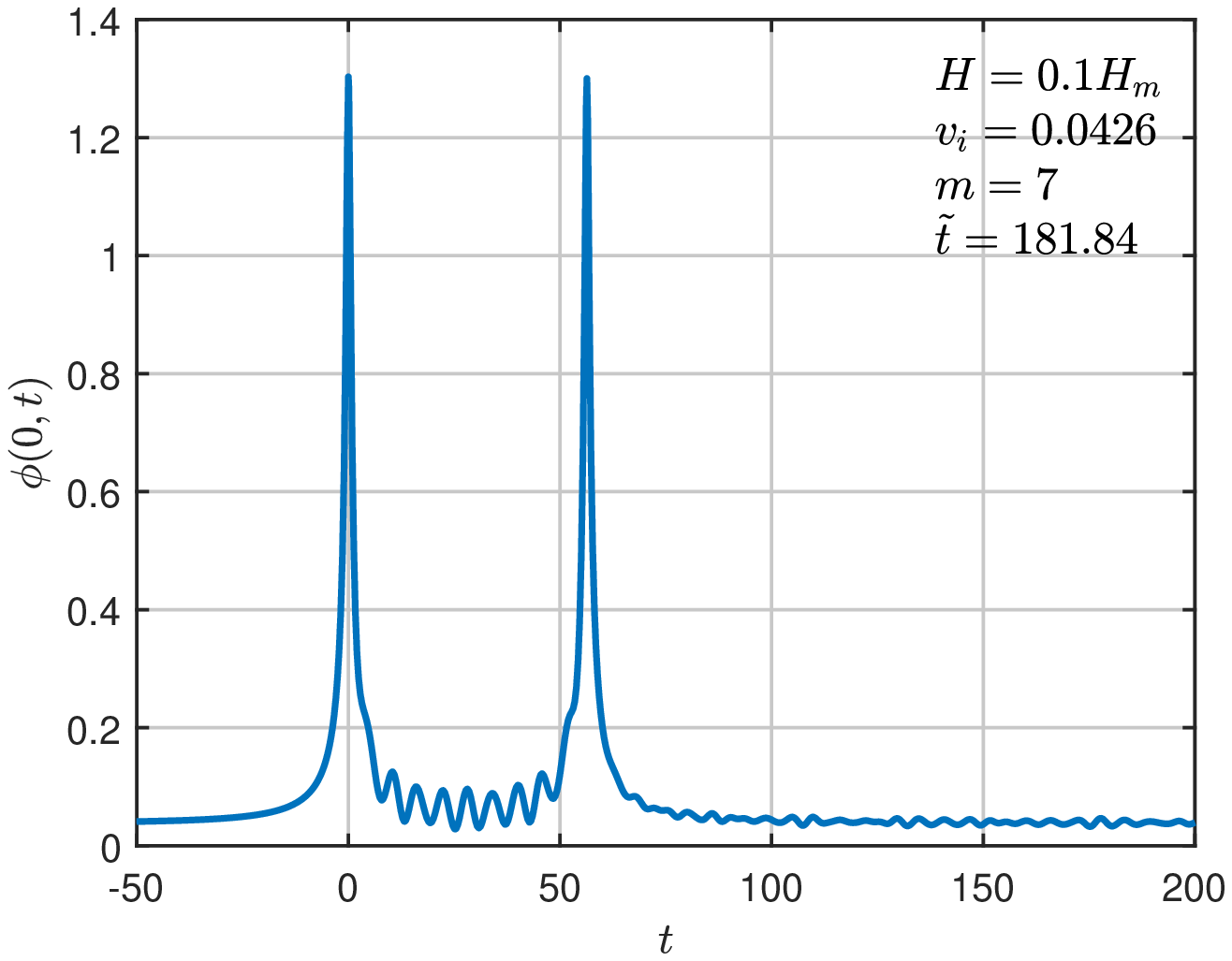}
\includegraphics[width=5.5cm]{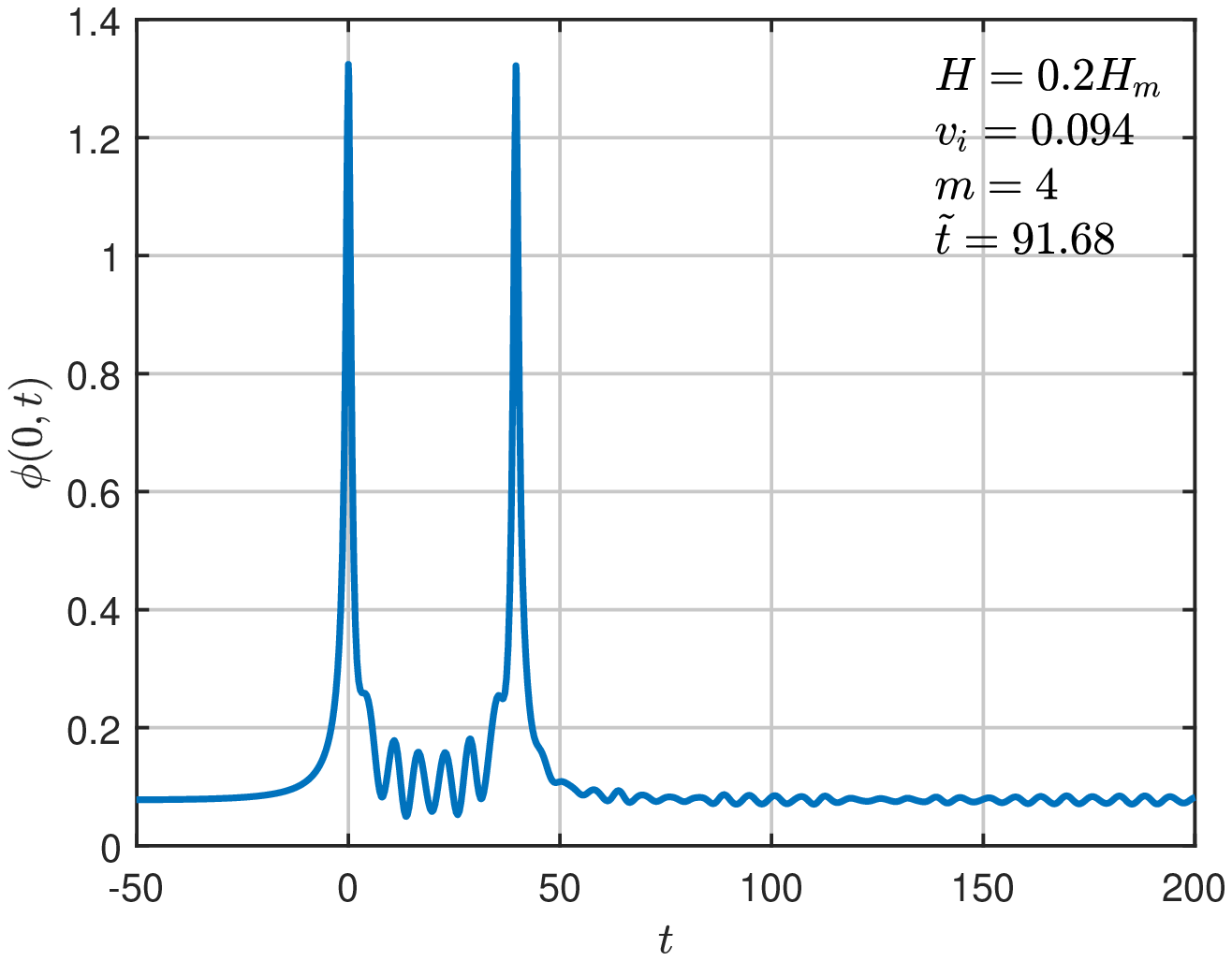}
\includegraphics[width=5.5cm]{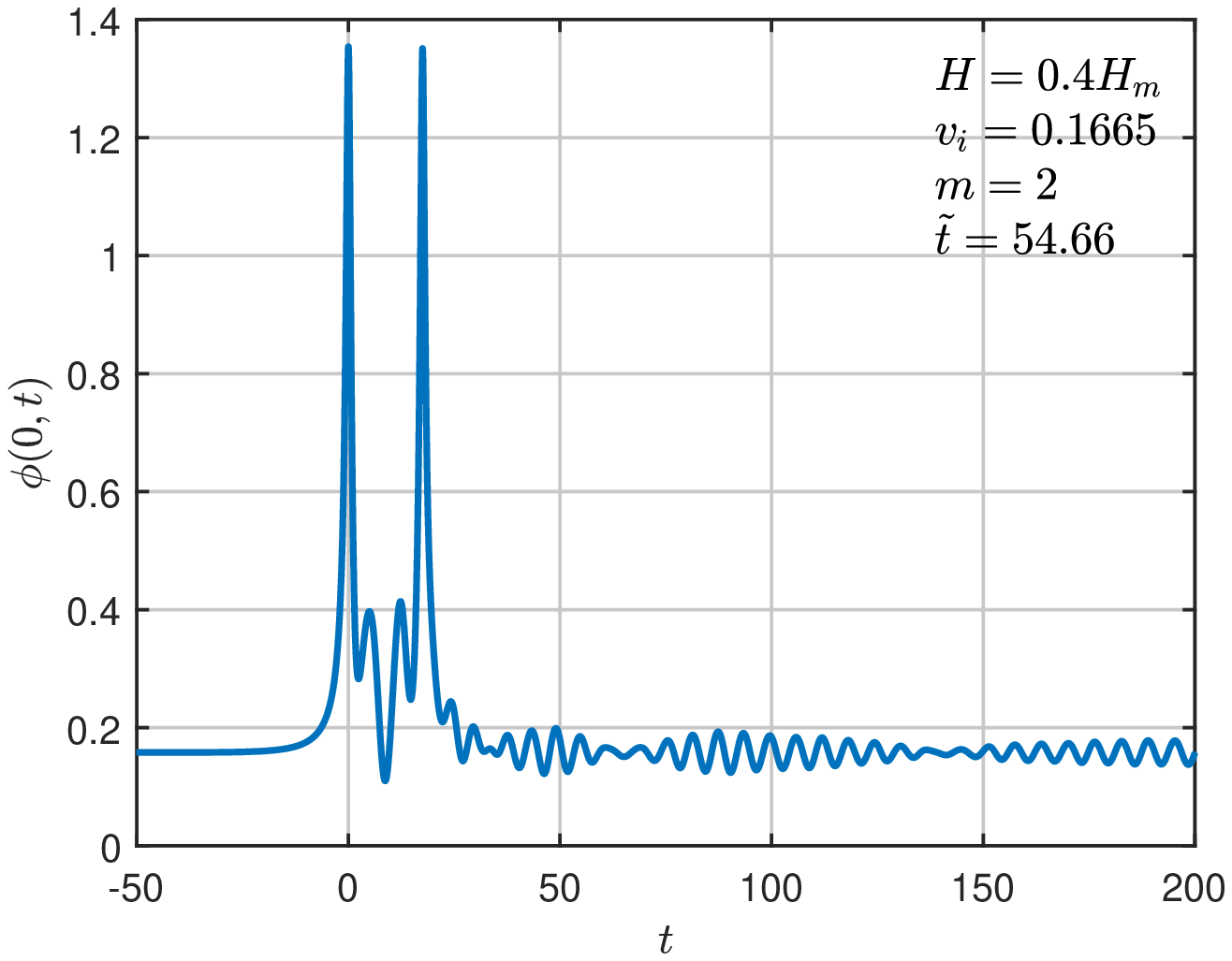}
\caption{$\Phi_{2+}(x)$ antikink scattering with $\phi_1$ boundary: scalar field at the boundary $\phi(0,t)$ as a funtion of $t$ for
collisions from the first visible two-bounce window.  a) $H=0.1H_m$,
b) $H=0.2H_m$, c) $H=0.4H_m$. The number $m$ of oscillations between
bounces for the first visible two-bounce windows is also indicated. In each figure the time $t$ was rescaled as $t-\tilde t$ such that $t=0$ corresponds to the first peak. }
\label{bounces}
\end{figure}

\begin{center}
\begin{figure}[H]
\begin{centering}
\includegraphics[width=14cm, height=6cm]{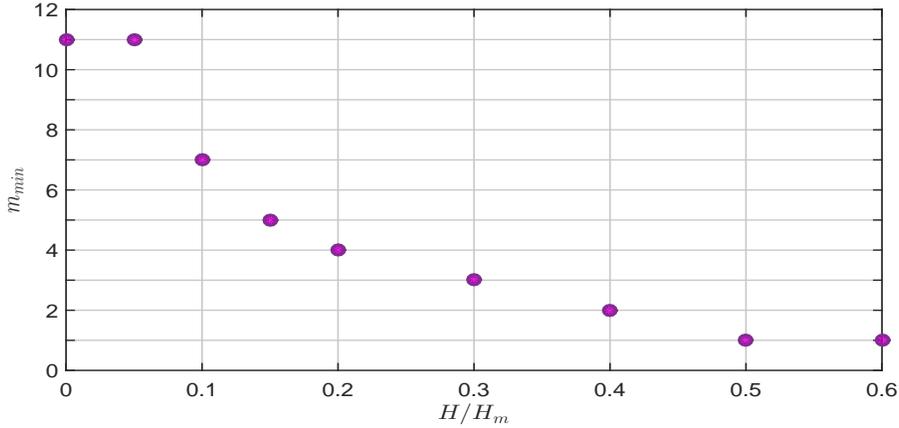}
\par\end{centering}
\caption{ $\Phi_{2+}(x)$ antikink scattering with $\phi_1$ boundary: order of first two-bounce windows as a function of $H/H_{m}$.}
\label{mVersusH}
\end{figure}
\par\end{center}

One remarkable effect of the variation of $H$ is indicated by the
integer $m$, the number of oscillations between bounces of $\phi(0,t)$
for the first visible two-bounce windows. This can be seen in the Figs.
\ref{bounces} and \ref{mVersusH}. For small values of $H$ ($H\leq0.1H_{m}$),
the structure of bounce windows is slightly altered, as can be seen
in Fig. \ref{bounces}a. The first two-bounce windows in this figure is
labeled $m=7$. This corresponds to the number of oscillations between
bounces shown in Fig. \ref{bounces}a. This is smaller than the value
$m=11$ obtained for $H=0$ (our $m$ corresponds to $n-1$ in the Ref. \cite{kk_nonintegr5}). Growing
more the parameter $H$, we see that $m$ decreases - compare Fig. \ref{bounces}b ($H=0.2H_m$, $m=4$) with Fig. \ref{bounces}c ($H=0.4H_m$, $m=2$). A quantitative view of this decay of $m$ with $H$ is presented in the Fig. \ref{mVersusH}. There one can see that $m$
decreases until the value $m=1$ of the first window for $H=0.5H_{m}$. This
shows that the analog magnetic field contributes to reduce the order
of the first two-bounce windows, approaching to those expected from CSW
mechanism. However, this occurs at the price of an increasing distortion
around the bounces of $\phi(0,t)$, as can be seen in Figs. \ref{bounces}b
and \ref{bounces}c. For $H>H_{c}\simeq 0.6H_{m}$ two-bounce windows disappear and the vibrational mode of the boundary now contributes to the production of kink during the scattering.

We have also checked that even in the regime of $H$ connected with an
irregular pattern of oscillations between bounces (see for instance Fig.
\ref{bounces}c for $H=0.4H_{m}$), the classification of $m$ used
here is in accord to the CSW mechanism. This was confirmed by the expected linear behavior between the time interval between bounces and the parameter $m$ for both regimes $H=0.1H_{m}$
(where the fluctuations are regular) and $H=0.4H_{m}$ (where the
fluctuations are irregular).

\section{Kink-Boundary}
Here we will consider scattering with a kink in the sector $(0,1)$ with the boundary. The kink is the static solution $\Phi_{1+}(x)$ given by the Eq. (\ref{Phi1+}), now boosted with a velocity $v_i$ with $x(t=0)=a$ and given by:
\be
\Phi_{1+}(v_i,x,t)=\sqrt{\frac{1+\tanh(\gamma(x-v_it-a))}{2}}.
\ee
The Fig. \ref{perfil-phi6} shows that this kink  can be connected only with solutions  $\tilde\phi_1(x)$ and $\tilde\phi_3(x)$. In the following we consider separately the two possibilities.

\subsection{Kink-$\tilde{\phi}_{1}$ boundary}

For this solution we consider $-H_{m}<H<0$. 
The initial conditions are
\begin{eqnarray}
\phi(x,0)&=&\Phi_{1+}(v,x,0)+\tilde\phi_{1}(x)-1,\\
\dot\phi(x,0)&=& \dot\Phi_{1+}(v,x,t)|_{t=0}.
\end{eqnarray}
with $ a=-12.50 $. 
\begin{figure}
	\includegraphics[scale=0.8]{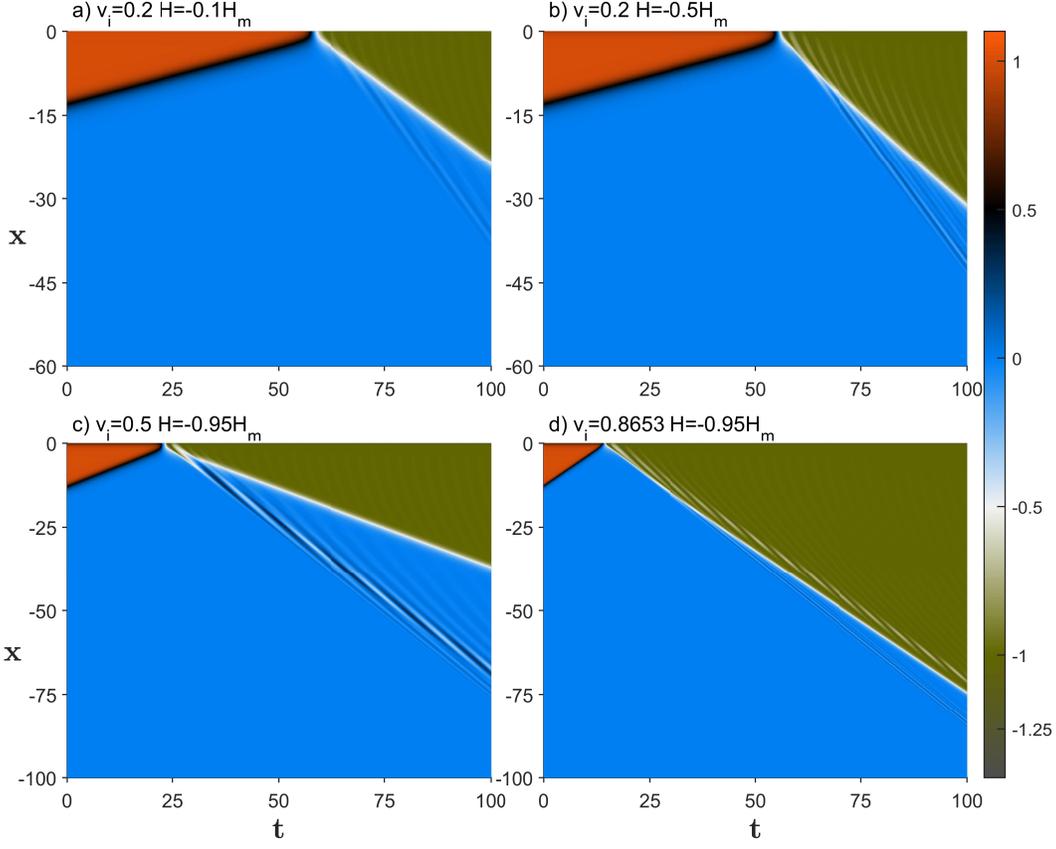} 
	\caption{$\Phi_{1+}$ kink scattering with $\tilde\phi_1$ boundary for different values of $v_i$ and $H$, showing: in a) and b), the changing of the topological sector with a produced $\Phi_{1-}$ antikink; in c) and d), a produced oscillon and $\Phi_{1-}$ antikink.}
	\label{3d2}
\end{figure}

\begin{figure}[h]
	\begin{centering}
		\includegraphics{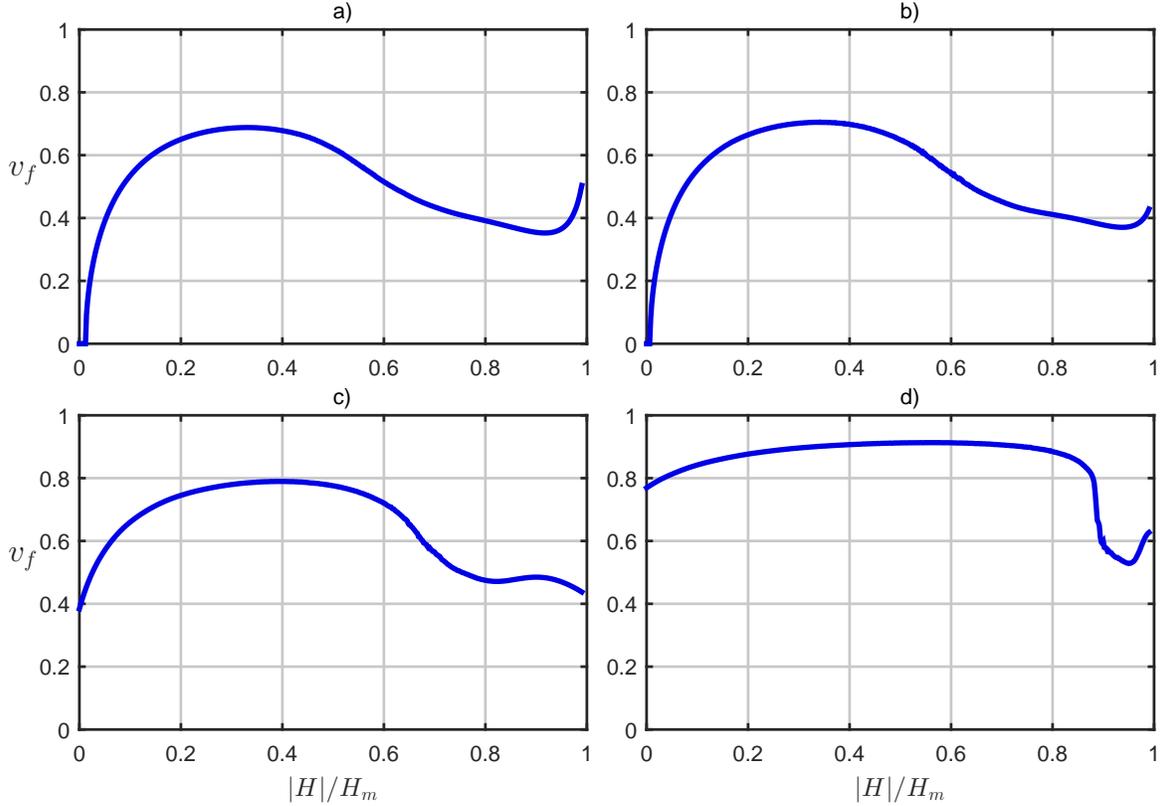} 
		\par\end{centering}
	\caption{Final velocity modulus $v_f$ of the produced $\Phi_{1-}$ antikink as a function of  $|H|/H_m$ with a) $v_i=0.05$ b) $v_i=0.20$ c) $v_i=0.50$  d) $v_i=0.80$. }
	\label{vfinal}
\end{figure}

The Fig. \ref{3d2} shows different aspects of the scattering process.  We found that, as a result of the scattering there is a changing of the topological sector with the production a $\Phi_{1-}$ antikink. Note also that the nature of the boundary $\tilde{\phi}_{1}$ has changed due to the scattering to another boundary $-\tilde\phi_3$. In this way we have the process $\Phi_{1+}+ \tilde\phi_1 \to  \Phi_{1-} + (-\tilde\phi_3)$. The final $-\tilde\phi_3$ boundary has no bound state, only a continuum of states.  The modulus $v_f$ of the velocity of the produced antikink has an intricate dependence with $v_i$ and $H$, as shown for instance in the Fig. \ref{vfinal}. Another interesting effect is the appearing of a travelling oscillon, as shown in the Fig. \ref{3d2}b. This effect starts to appear for $|H|\approx 0.70H_m$ but it is more evident for  $|H|>0.90H_m$. The oscillon has the scalar field around the vacuum   $ \phi=0 $ when its velocity is larger than $v_f$, as in the example of the Fig.\ref{3d2}c. If its velocity is lower than $v_f$,  the scalar field oscillating around the vacuum   $ \phi=-1$. Note also that, contrarily to the rich pattern observed for antikink-boundary scattering in the Fig. \ref{mosaico}, here we have that for all phase space  $(v_i,H)$ the final state of the boundary is close to the vacuum $\phi = -1$.

\subsection{Kink-$\tilde{\phi}_{3}$ boundary}
For this solution we have $0\le H<\infty$. However, to ease the comparison with the other cases, in our simulations we will consider the specific range $0 \le H \le H_m$. The initial conditions are
\begin{eqnarray}
\phi(x,0)&=&\Phi_{1+}(v,x,0)+\tilde\phi_{3}(x)-1,\\
\dot\phi(x,0)&=& \dot\Phi_{1+}(v,x,t)|_{t=0}.
\end{eqnarray}
with $ a=-12.50 $. 

\begin{figure}
	\begin{centering}
		\includegraphics[scale=0.8]{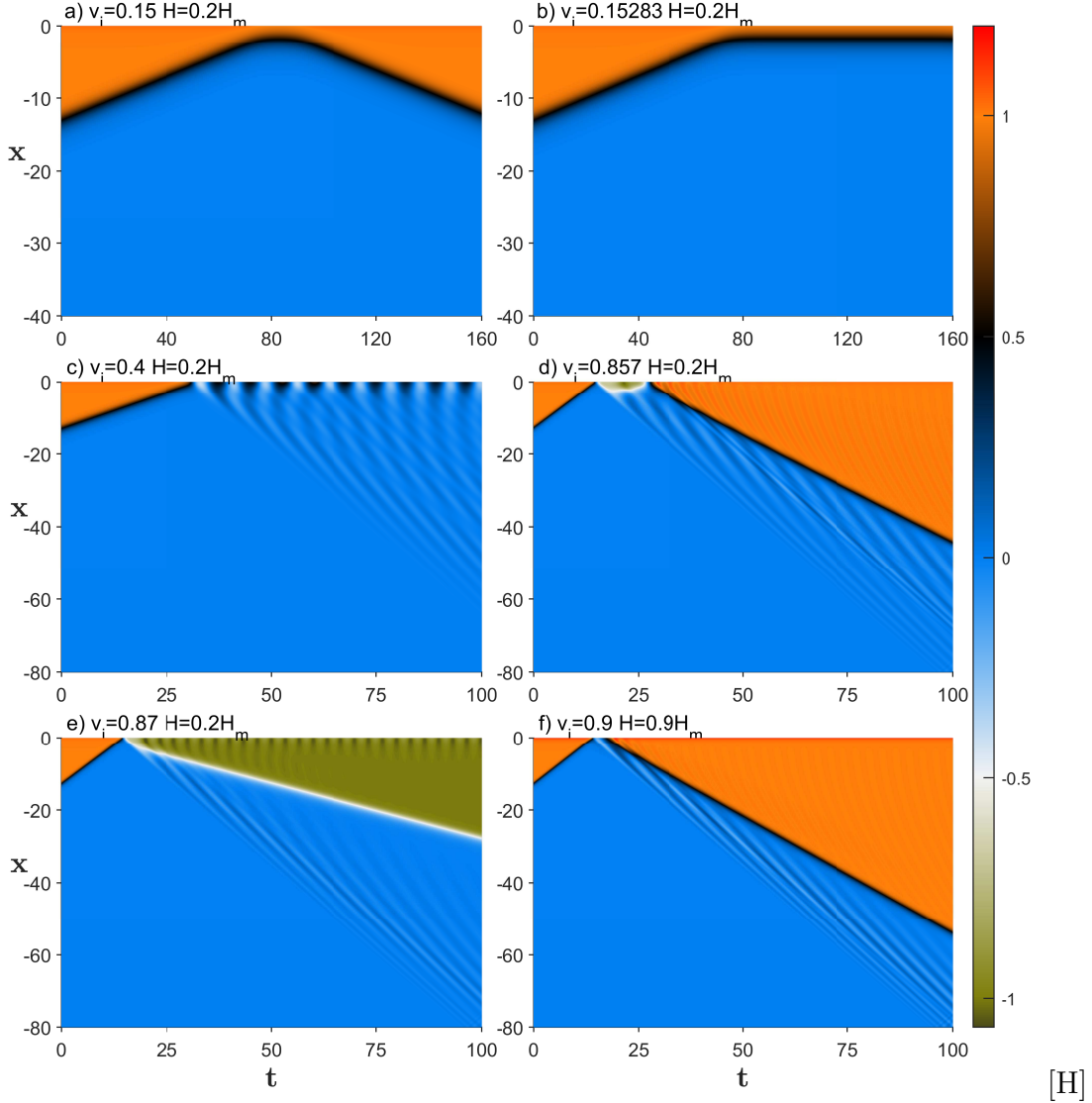}[H] 
		\par\end{centering}
	\caption{$\Phi_{1+}$ kink scattering with $\tilde\phi_3$ boundary, showing a) quasi-elastic return b) kink with energy to reach the energy equilibrium point c) emission of radiation by the unstable $\phi_2$ boundary d) Typical collision on top orange region of Fig. \ref{mosaico2} e) Inelastic scattering f) Final state for large $H$.  
\label{boundary2}}
\end{figure}

\begin{figure}[H]
	\includegraphics[scale=0.9]{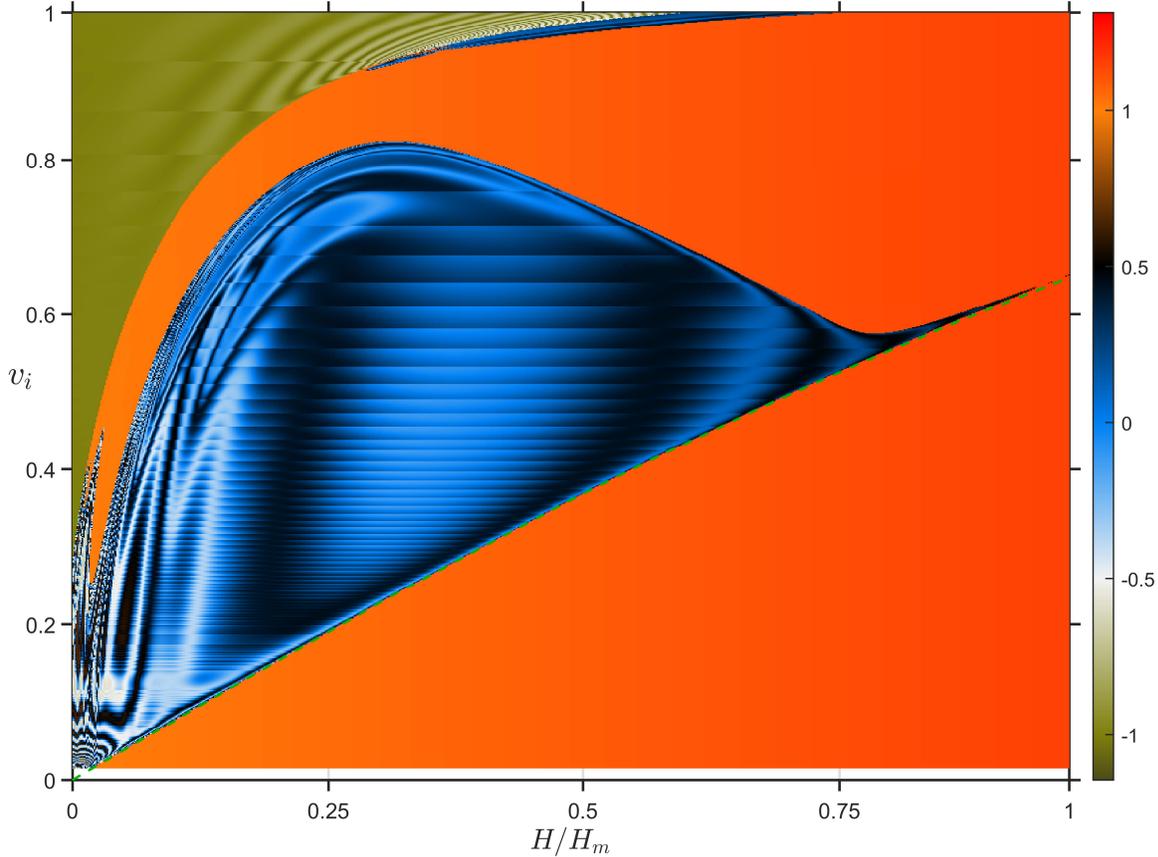}
	\caption{Final state of the scalar field $\phi\left( 0,t_{f}\right)$, with $t_{f}$ being the nearest integer of $\frac{|a|}{v_{i}}+100$.}
	\label{mosaico2}
\end{figure}

In this case the boundary presents an initially-repulsive force. The Fig. \ref{boundary2} shows the pattern of scattering for different values of $v_i$ and $H$. The Fig. \ref{mosaico2} shows the value of the scalar field at the boundary for a large scale of time. For small initial velocities, the kink do not even reach the boundary at $x=0$, and the scattering is almost elastic, with a small emission of radiation (see the Fig. \ref{boundary2}a).
Increasing the value of $v_i$ until a certain velocity $v_*$, the kink becomes stuck to the boundary, with very low rate of emitted radiation, as shown in the Fig. \ref{boundary2}b. This final state is the unstable configuration $\phi_2$. This is supported by energy considerations. Indeed, since there is a small amount of emitted radiation, we can obtain $v_*$ considering the conservation of energy between initial and final states. For the initial state we have $ E_i= \frac{1}{4}\gamma(v_{*}) + E[\tilde{\phi}_{3}]$ whereas for the final state we have  $ E_f=E[\phi_2] $. The equality $ E_i=E_f $ gives (see a similar reasoning for the $\phi^4$ model in the Ref. \cite{dorey1})
\begin{equation}
\label{v*}
v_{*}(H)=\sqrt{1-\left[3\phi_{2}\left(0\right)^{4}+3\tilde{\phi}_{3}\left(0\right)^{4}-2\phi_{2}\left(0\right)^{2}-2\tilde{\phi}_{3}\left(0\right)^{2}-1\right]^{-2}}.
\end{equation}
The curve $ v_{*}(H) $ is included in the Fig. (\ref{mosaico2}) (green traced line). Note the agreement between this figure, which is constructed with the results of numerical solutions of the kink-boundary scattering and the expression given by the Eq. (\ref{v*}).  Increasing the initial velocity above $v_{*}$, there is an intermediate region where we observe that the kink approaches, overcome the potential barrier and is absorbed by the boundary, that now radiates in a typical pattern of the unstable solution $\phi_2$ (see the Fig. \ref{boundary2}c and the discussion of the end of Sect. \ref{half}).  Here one has a process like $\Phi_{1+}+\tilde\phi_{3} \to \phi_{2}$.  This type of scattering corresponds to the central blue region of the Fig. \ref{mosaico2}. Keeping $H$ fixed and increasing $v_i$, there occurs a scenario like that of the Fig. \ref{boundary2}d where the nature of the $\tilde\phi_{3}$ boundary does not change due to the scattering. There is the emission of a  $\Phi_{1+}$ kink, followed by the emission of radiation from the continuum states of the boundary. This type of scattering corresponds to the upper orange region of the Fig. \ref{mosaico2}. 
 Increasing further the velocity, one has the scenario of the Fig.  \ref{boundary2}e, which shows the changing of the topological sector and the  production of an $\Phi_{1-}$ antikink. This corresponds to the green region of the Fig.\ref{mosaico2}.The possibility of changing the topological sector decreases for large values of $H$.  Larger values of $H$ in the same orange region of  Fig. \ref{mosaico2} we have the effect that during the collision the scalar field at the center of mass oscillates around $\phi=0$, as in the Fig. \ref{boundary2}f.

\section{Conclusions}
In this work we have considered the $\phi^6$ model with a Neumann boundary condition characterized by the parameter $H$. We found several solutions for kink, antikink and boundary. Stability analysis for the antikink and kink lead to a translational mode. A similar procedure for the boundary solutions gives the behavior of the  squared frequency of discrete modes as a function of $H$. This resulted in three regular and stable solutions for the boundaries. The study of the unstable boundary was important since the nature of a boundary can be changed to an unstable one due to the scattering process. The changing of the nature of a regular and stable boundary due to the scattering was verified in some examples in two ways: i) a topological argument - the changing of the topological sector, and ii) energy conservation leading to an accurate description of the critical velocity as a function of $H$. After a period of time without any sensible alteration, and depending on the parameter $H$, there are two possibilities for the time evolution of an isolated unstable boundary: i) radiation being emitted continuously, or  ii) creation of a kink and decaying to a stable boundary. As initial conditions, we considered kink/antikink scattering with regular and stable boundaries.

Our numerical analysis showed that the antikink-boundary scattering 
model has a rich pattern. We report the following phenomena, depending
on the initial velocity and the parameter $H$: i) the production
of a new kink from the boundary at $x=0$ that travels with the reflected
antikink. Such reflected antikink-kink pair can have a finite number
of mutual collisions separating afterwards. Other possibility is the
pair travelling together and colliding an infinite number of times, in a bion-like
state. Such effect was observed previously in the $\phi^{4}$ model
(see the Ref. \cite{dorey1}). ii) the presence of a critical escape
velocity, above which the antikink always receedes to $-\infty$ without
the production of a new travelling kink at the boundary. iii) the
formation of two-bounce windows.

For the kink-boundary scattering here are two possibilities for a regular and stable boundary. For one boundary the scalar field always changes to the other topological sector. As a result, the new boundary emits an antikink with or without an oscillon. Another possibility for the initial boundary has a richer pattern. In this case the interaction with the kink is repulsive. This means that there is a critical velocity where the kink become trapped by the boundary. Depending on the parameters $(v_i,H)$ one can also see the changing of
 the boundary with the emission of radiation  or the changing of the topological sector.

When compared with the results for the $\phi^4$ model \cite{dorey1}, despite its higher nonlinearity, the structure revealed by the antikink-boundary collisions for the $\phi^6$ model is simpler. Indeed, as we saw in Fig.  \ref{vcversusH}, $v_c$ grows monothonically with $H$ for $0<H<H_m$, favoring the emission of a kink from the boundary. On the contrary, for the $\phi^4$ model and $0<H\lesssim 0.5H_m$, $v_c$ initially decreases with $H$, favoring the 1-bounce region. Fig.  \ref{mosaico} shows that the region close to  $H \sim H_m$ for the $\phi^6$ model has a clear separation between two regions: 1-bounce (blue) and kink formation from the boundary (red). The corresponding region for the $\phi^4$ model (see the Fig. 3 from Ref. \cite{dorey1}) is intricate, and the $v_c$ has no functional form.  The complexity of the $\phi^6$ model comes from the existence of two topological sectors. Even in one topological sector there are more solutions for kink, antikink and boundaries. This results, for the $\phi^6$ model,  in different structures for antikink-boundary and kink-boundary scattering and in the possibility of changing the topological sector. 

Despite our investigation being mostly numerical, we connected some of the reported effects to the presence of bound states and continuum states in the system antikink-boundary or kink-boundary. Also, our results extended, for $H\neq0$, the known results for scattering in the full-line.


\section{Acknowledgements}
F.C.L, F.C.S., K.Z.N and A.R.G thank FAPEMA - Funda\c c\~ao de Amparo \`a Pesquisa e ao Desenvolvimento do Maranh\~ao through grants PRONEX $01452/14$, PRONEM $01852/14$, Universal $01061/17$, $01191/16$, $01332/17$, $01441/18$ and $BD-00128/17$. A.R.G thanks CNPq (brazilian agency) through grants  $437923/2018$-5 and $311501/2018$-4 for financial support. This study was financed in part by the Coordena\c c\~ao de Aperfei\c coamento de Pessoal de N\' ivel Superior - Brasil (CAPES) - Finance Code 001. Gomes thanks R. Casana for discussions. The authors thank T. Roma\'nczukiewicz, P. Dorey and Ya. Shnir for the interesting comments about scattering with other boundaries and the existence of internal modes of the boundary solution.  


\end{document}